\documentclass[preprint,prd,aps,showpacs,showkeys,nofootinbib]{revtex4}
\usepackage{graphicx}
\usepackage{dcolumn}
\usepackage{bm}
\usepackage{diagbox}
\usepackage{color}
\usepackage[dvipsnames]{xcolor}
\usepackage{amssymb}
\usepackage{makecell}
\usepackage{subfigure}
\usepackage{amsmath}
\usepackage{epstopdf}

\definecolor{light-gray}{gray}{0.88}
\definecolor{dark-gray}{gray}{0.40}

\def\pds{\left| p_{D^\ast}\right|^2}
\def\mds{M_{D^\ast}^2}
\def\cvl{C_{VL}^\ell}
\def\cvr{C_{VR}^\ell}

\def\call{C_{AL}^\ell}

\def\cplconj{C_{PL}^{\ell\ast}}

\def\cvlsq{\left|C_{VL}^\ell\right|^2}

\def\calsq{\left|C_{AL}^\ell\right|^2}

\def\cplsq{\left|C_{PL}^\ell\right|^2}

\def\azsq{A_0^2}
\def\aosq{A_1^2}
\def\atsq{A_2^2}
\def\vsq{V^2}
\def\mbmd{\left(M_B + M_{D^\ast}\right)}
\def\mbmc{\left(m_b + m_c\right)}
\def\mbmdq2{\left(M_B^2 - M_{D^\ast}^2 - q^2\right)}
\def\re{\mathcal{R}}
\def\nn{\nonumber}

\def\ctlconj{C_{TL}^{\ell\ast}}

\def\ctrconj{C_{TR}^{\ell\ast}}
\def\ctlsq{\left|C_{TL}^\ell\right|^2}
\def\ctrsq{\left|C_{TR}^\ell\right|^2}
\def\mbmdsq{\left(M_B^2 - M_{D^\ast}^2 \right)}
\def\mbmdqsq{\left(M_B^2 - M_{D^\ast}^2 - q^2\right)}
\def\mbmdqthsq{\left(M_B^2 + 3 M_{D^\ast}^2 - q^2\right)}
\def\cvr{C_{VR}^\ell}
\def\car{C_{AR}^\ell}

\def\ctrconj{C_{TR}^{\ell\ast}}
\def\mB{M_B}
\def\mDs{M_{D^\ast}}
\def\pDmag{|p_{D^\ast}|}
\def\ctrsq{\left|C_{TR}^\ell\right|^2}
\def\Dst{{D^*}}

\begin{document}

\title{Theoretical Corrections of $R_D$ and $R_{D^*}$}
\author{Xin-Xin Long$^{1,2,3}$, Shu-Min Zhao$^{1,2,3}$\footnote{zhaosm@hbu.edu.cn}, Ming-Yue Liu$^{1,2,3}$, Xi Wang$^{1,2,3}$, Yi-Tong Wang$^{1,2,3}$, Zhong-Jun Yang$^{4}$, Xing-Xing Dong$^{1,2,3}$, Tai-Fu Feng$^{1,2,3,4}$}

\affiliation{$^1$ Department of Physics, Hebei University, Baoding 071002, China}
\affiliation{$^2$ Key Laboratory of High-precision Computation and Application of Quantum Field Theory of Hebei Province, Baoding 071002, China}
\affiliation{$^3$ Research Center for Computational Physics of Hebei Province, Baoding 071002, China}
\affiliation{$^4$ Department of Physics, Chongqing University, Chongqing 401331, China}
\date{\today}

\begin{abstract}
$R_{D^{(*)}}$ is the ratio of branching ratio $\overline{B} \rightarrow D^{(*)}\tau\overline{\nu}_{\tau}$ to $\overline{B} \rightarrow D^{(*)}l\overline{\nu}_{l}~(l=e,~\mu)$. There is a gap of 2$\sigma_{exp}$ or more between its experimental value and the prediction under the standard model(SM). People extend the MSSM with the local gauge group $U(1)_X$ to obtain the $U(1)_X$SSM. Compared with MSSM, $U(1)_X$SSM has more superfields and effects. In $U(1)_X$SSM, we research the semileptonic decays $\overline{B} \rightarrow D^{(*)}l\overline{\nu}_{l}$ and calculate $R_{D^{(*)}}$.
The numerical results of $R_{D^{(*)}}$ are further corrected under $U(1)_X$SSM, which are much better
than the SM predictions. After correction, the theoretical value of $R_{D^{(*)}}$ can reach
 one $\sigma_{exp}$ range of the averaged experiment central value.
\end{abstract}


\maketitle

\section{Introduction}
To date, the standard model (SM) is one of the most successful models in particle physics. It gives the ideal theoretical values of many experiments, but there are exceptions. Some experiments involving B meson decay have shown deviations from the SM expectation. The most statistically significant deviation is seen in the combination of $R_{D^{(*)}}$(the ratios of the branching fraction of $\overline{B} \rightarrow D^{(*)}\tau\overline{\nu}_{\tau}$ to that of
$\overline{B} \rightarrow D^{(*)}l\overline{\nu}_{l}, l=e, \mu$).
 Babar~\cite{1,2}, Belle~\cite{3,4,5}, and LHCb~\cite{6} collaborate on measurements of $R_{D^{(*)}}$. A number of people also carry out the calculation of the theoretical value under SM~\cite{7,8,9,10,11,12}, and the averaged value is synthesized by Ref. \cite{13}. Therefore, we give the theoretical and experimental values as follows :
\begin{eqnarray}
&&R_D^{exp}=0.356 \pm 0.029_{total},~~~~~~~~R_{D^*}^{exp} =0.284 \pm 0.013_{total},\label{tnph1}\nonumber\\
&&R_D^{SM}=0.298\pm 0.004,~~~~~~~~~~~~R_{D^*}^{SM}=0.254\pm 0.005.\label{RDSM}
\end{eqnarray}
The difference between the theoretical central value of ${R_D}$ and the experimental central value is $2\sigma_{exp}$, and the difference between the theoretical central value of  $R_{D^*}$ and the experimental central value is $2.3\sigma_{exp}$.

Due to the deviation between the experimental value and the expected value
  of the SM, people turn their attention to new physics (NP) to further correct the theoretical results. Many studies \cite{R01,R02,R3,R4,R5,R6,R7,R8,R9,R10,R11,R12,R13,R14,R15,R16,R17,R18,R19,R20,R21} have been carried out in order to obtain the NP explanation of the deviation. The minimal supersymmetric extension of the standard model (MSSM)~\cite{MSSM} is a well-known extension that has been used in many decay processes. However, when interpreting neutrino oscillations, we need a tiny neutrino mass that MSSM cannot naturally produce. In order to solve this difficulty, people extend it again on the basis of MSSM. In the past work\cite{BLM},
 we know that in the minimal supersymmetric extension of the SM with local gauged B and L (BLMSSM)
 the theoretical value has been improved. However, the improved extent is small and can not well compensate the gap between SM prediction and experiment value. $U(1)_X$SSM \cite{UX1,UX2,UX3} is an extended model based on MSSM, and the local gauge group is $SU(3)_C\otimes SU(2)_L \otimes U(1)_Y \otimes U(1)_X$. Compared with BLMSSM, $U(1)_X$SSM has gauge mixing terms and gauge mixing constant, and they produce new effects.
 Compared with MSSM, the model obtained by $U(1)_X$ extension has more superfields.
 Among these superfields, the right-handed neutrinos and the added Higgs singlets solve the problem of tiny neutrino mass. Due to the existence of $\mu\hat{H}_u\hat{H}_d$ and $\lambda_H\hat{S}\hat{H}_u\hat{H}_d$ in $U(1)_X$SSM, it has the effective $\mu_{H}'=\mu+\lambda_Hv_S/\sqrt{2}$. Thus, the problem of $\mu$ of MSSM is alleviated.

This paper discusses the theoretical correction from NP, and the advantages of $U(1)_X$SSM help us to achieve this goal. Next, this paper will first introduce the superfields and the general soft breaking terms of $U(1)_X$SSM and the related couplings in Sec.II. Then, the formulas of $R_{D^{(*)}}$ in detail are given in Sec.III. After introducing the formulas, we give the Feynman diagrams and part of the calculation process. Subsequently, we visualize the data through the software Mathematica and display it in the form of pictures in Sec.IV. Finally, this paper is summarized in Sec.V. Some formulas are shown in the Appendix.

\section{The $U(1)_X$SSM}
$U(1)_X$SSM local gauge group is $SU(3)_C\otimes SU(2)_L \otimes U(1)_Y\otimes U(1)_X$. As an extension of MSSM, it introduces three Higgs singlets $\hat{\eta},~\hat{\bar{\eta}},~\hat{S}$ and three-generation right-handed neutrinos $\hat{\nu}_i$. Compared with MSSM, the advantages of $U(1)_X$SSM are:

1. Through the see-saw mechanism, the light neutrinos obtain tiny mass at the tree level.

2. The neutral CP-even parts of the introduced  $\hat{\eta},~\hat{\bar{\eta}},~\hat{S}$, $H_u$ and $H_d$ are mixed together to form a $5\times5$ mass squared matrix. The lightest CP-even Higgs mass can be improved at tree level.\\
The particle contents and charge assignments can be found in Refs. \cite{model2}.

The superpotential of this model is
\begin{eqnarray}
&&W=l_W\hat{S}+\mu\hat{H}_u\hat{H}_d+M_S\hat{S}\hat{S}-Y_d\hat{d}\hat{q}\hat{H}_d-Y_e\hat{e}\hat{l}\hat{H}_d+\lambda_H\hat{S}\hat{H}_u\hat{H}_d
\nonumber\\&&~~~~~~+\lambda_C\hat{S}\hat{\eta}\hat{\bar{\eta}}+\frac{\kappa}{3}\hat{S}\hat{S}\hat{S}+Y_u\hat{u}\hat{q}\hat{H}_u+Y_X\hat{\nu}\hat{\bar{\eta}}\hat{\nu}
+Y_\nu\hat{\nu}\hat{l}\hat{H}_u.
\end{eqnarray}

We use $v_{u},~v_{d},~v_\eta$,~ $v_{\bar\eta}$ and $v_S$ to represent the vacuum expectation values (VEVs) of the Higgs superfields $H_u$, $H_d$, $\eta$, $\bar{\eta}$ and $S$. Then, we show the explicit forms of two Higgs doublets and three Higgs singlets here
\begin{eqnarray}
&&H_{u}=\left(\begin{array}{c}H_{u}^+\\{\frac{1}{\sqrt{2}}}\Big(v_{u}+H_{u}^0+iP_{u}^0\Big)\end{array}\right),
~~~~~~
H_{d}=\left(\begin{array}{c}{\frac{1}{\sqrt{2}}}\Big(v_{d}+H_{d}^0+iP_{d}^0\Big)\\H_{d}^-\end{array}\right),
\nonumber\\
&&\eta={\frac{1}{\sqrt{2}}}\Big(v_{\eta}+\phi_{\eta}^0+iP_{\eta}^0\Big),~~~~~~~~~~~~~~~
\bar{\eta}={\frac{1}{\sqrt{2}}}\Big(v_{\bar{\eta}}+\phi_{\bar{\eta}}^0+iP_{\bar{\eta}}^0\Big),\nonumber\\&&
\hspace{4.0cm}S={\frac{1}{\sqrt{2}}}\Big(v_{S}+\phi_{S}^0+iP_{S}^0\Big).
\end{eqnarray}
And two angles are defined as $\tan\beta=v_{u}/v_{d}$ and $\tan\beta_\eta=v_{\bar{\eta}}/v_{\eta}$.

The soft SUSY breaking terms of this model are shown as
\begin{eqnarray}
&&\mathcal{L}_{soft}=\mathcal{L}_{soft}^{MSSM}-B_SS^2-L_SS-\frac{T_\kappa}{3}S^3-T_{\lambda_C}S\eta\bar{\eta}
+\epsilon_{ij}T_{\lambda_H}SH_d^iH_u^j\nonumber\\&&\hspace{1.5cm}
-T_X^{IJ}\bar{\eta}\tilde{\nu}_R^{*I}\tilde{\nu}_R^{*J}
+\epsilon_{ij}T^{IJ}_{\nu}H_u^i\tilde{\nu}_R^{I*}\tilde{l}_j^J
-m_{\eta}^2|\eta|^2-m_{\bar{\eta}}^2|\bar{\eta}|^2-m_S^2S^2\nonumber\\&&\hspace{1.5cm}
-(m_{\tilde{\nu}_R}^2)^{IJ}\tilde{\nu}_R^{I*}\tilde{\nu}_R^{J}
-\frac{1}{2}\Big(M_X\lambda^2_{\tilde{X}}+2M_{BB^\prime}\lambda_{\tilde{B}}\lambda_{\tilde{X}}\Big)+h.c~~.
\end{eqnarray}

There are two Abelian groups $U(1)_Y$ and $U(1)_X$ simultaneously in $U(1)_X$SSM, resulting in a new effect that does not exist in MSSM: the gauge kinetic mixing. This effect can also be induced through RGEs even with zero at $M_{GUT}$.

In general, the covariant derivatives of $U(1)_X$SSM can be written as~\cite{UX4,BL1,BL2,Gmass}
\begin{eqnarray}
&&D_\mu=\partial_\mu-i\left(\begin{array}{cc}Y,&X\end{array}\right)
\left(\begin{array}{cc}g_{Y},&g{'}_{{YX}}\\g{'}_{{XY}},&g{'}_{{X}}\end{array}\right)
\left(\begin{array}{c}A_{\mu}^{\prime Y} \\ A_{\mu}^{\prime X}\end{array}\right)\;.
\end{eqnarray}
$A_{\mu}^{\prime Y}$ and $A^{\prime X}_\mu$ represent the gauge fields of $U(1)_Y$ and $U(1)_X$, respectively. Under the condition that the two Abelian gauge groups are not broken, we change the basis of the above equation by rotation matrix $R$~\cite{UX4,BL2}. As follows
\begin{eqnarray}
&&\left(\begin{array}{cc}g_{Y},&g{'}_{{YX}}\\g{'}_{{XY}},&g{'}_{{X}}\end{array}\right)
R^T=\left(\begin{array}{cc}g_{1},&g_{{YX}}\\0,&g_{{X}}\end{array}\right).
\end{eqnarray}

Three neutral gauge bosons $A^{X}_\mu,~A^Y_\mu$ and $V^3_\mu$ mix together at the tree level, whose mass matrix
is shown in the basis $(A^Y_\mu, V^3_\mu, A^{X}_\mu)$
\begin{eqnarray}
&&\left(\begin{array}{*{20}{c}}
\frac{1}{8}g_{1}^2 v^2 &~~~ -\frac{1}{8}g_{1}g_{2} v^2 & ~~~\frac{1}{8}g_{1}(g_{{YX}}+g_{X}) v^2 \\
-\frac{1}{8}g_{1}g_{2} v^2 &~~~ \frac{1}{8}g_{2}^2 v^2 & ~~~~-\frac{1}{8}g_{2}(g_X+g_{{YX}}) v^2\\
\frac{1}{8}g_{1}(g_{{YX}}+g_{X}) v^2 &~~~ -\frac{1}{8}g_{2}(g_{{YX}}+g_{X}) v^2 &~~~~ \frac{1}{8}(g_{{YX}}+g_{X})^2 v^2+\frac{1}{8}g_{{X}}^2 \xi^2
\end{array}\right),\label{gauge matrix}
\end{eqnarray}
with $v^2=v_{u}^2+v_{d}^2$ and $\xi^2=v_\eta^2+v_{\bar{\eta}}^2$.

Through the Weinberg angle $\theta_{W}$ and the new mixing angle $\theta_{W}'$, we can diagonalize the mass matrix in Eq.(\ref{gauge matrix}). The new mixing angle $\theta_{W}'$ is defined by the following formula
\begin{eqnarray}
\sin^2\theta_{W}'\!=\!\frac{1}{2}\!-\!\frac{[(g_{{YX}}+g_{X})^2-g_{1}^2-g_{2}^2]v^2+
4g_{X}^2\xi^2}{2\sqrt{[(g_{{YX}}+g_{X})^2+g_{1}^2+g_{2}^2]^2v^4\!+\!8g_{X}^2[(g_{{YX}}+g_{X})^2\!-\!g_{1}^2\!-\!g_{2}^2]v^2\xi^2\!+\!16g_{X}^4\xi^4}}.
\end{eqnarray}
The exact eigenvalues of Eq.(\ref{gauge matrix}) are deduced
\begin{eqnarray}
&&m_\gamma^2=0,\nonumber\\
&&m_{Z,{Z^{'}}}^2=\frac{1}{8}\Big([g_{1}^2+g_2^2+(g_{{YX}}+g_{X})^2]v^2+4g_{X}^2\xi^2 \nonumber\\
&&\hspace{1.1cm}\mp\sqrt{[g_{1}^2+g_{2}^2+(g_{{YX}}+g_{X})^2]^2v^4\!+\!8[(g_{{YX}}+g_{X})^2\!-\!g_{1}^2\!-\!
g_{2}^2]g_{X}^2v^2\xi^2\!+\!16g_{X}^4\xi^4}\Big).
\end{eqnarray}

The mass matrices can be found in the works~\cite{UU1,20}. Here, we show some couplings that need to be used later in this model.

1. The vertexes of $\bar{l}_i-\chi_j^--\tilde{\nu}^R_k(\tilde{\nu}^I_k)$.
\begin{eqnarray}
&&\mathcal{L}_{\bar{l}\chi^-\tilde{\nu}^R}=\frac{1}{\sqrt{2}}\bar{l}_i\Big\{U^*_{j2}Z^{R*}_{ki}Y_l^iP_L
-g_2V_{j1}Z^{R*}_{ki}P_R\Big\}\chi_j^-\tilde{\nu}^R_k,
\\&&\mathcal{L}_{\bar{l}\chi^-\tilde{\nu}^I}=\frac{i}{\sqrt{2}}\bar{l}_i\Big\{U^*_{j2}Z^{I*}_{ki}Y_l^iP_L
-g_2V_{j1}Z^{I*}_{ki}P_R\Big\}\chi_j^-\tilde{\nu}^I_k.
\end{eqnarray}

2. The vertexes of $\bar{\chi}_i^0-\nu_i-\tilde{\nu}^R_k(\tilde{\nu}^I_k)$.
\begin{eqnarray}
&&\mathcal{L}_{\bar{\chi}^0\nu\tilde{\nu}^R}=\frac{1}{2}\bar{\chi}_i^0\Big\{(-g_2N^*_{i2}+g_{YX}N^*_{i5}+g_1N^*_{i1})
\sum_{a=1}^3Z^{R*}_{ka}U_{ja}^{V*}P_L\nonumber\\&&\hspace{1.7cm}+
(-g_2N_{i2}+g_{YX}N_{i5}+g_1N_{i1})\sum_{a=1}^3Z^{R*}_{ka}U_{ja}^{V}P_R\Big\}\nu_j\tilde{\nu}^R_k,
\\&&\mathcal{L}_{\bar{\chi}^0\nu\tilde{\nu}^I}=-\frac{i}{2}\bar{\chi}_i^0\Big\{(-g_2N^*_{i2}+g_{YX}N^*_{i5}+g_1N^*_{i1})
\sum_{a=1}^3Z^{I*}_{ka}U_{ja}^{V*}P_L\nonumber\\&&\hspace{1.7cm}+
(g_2N_{i2}-g_{YX}N_{i5}-g_1N_{i1})\sum_{a=1}^3Z^{I*}_{ka}U_{ja}^{V}P_R\Big\}\nu_j\tilde{\nu}^I_k.
\end{eqnarray}

3. The vertexs of neutrino-slepton-chargino and neutralino-lepton-slepton.
\begin{eqnarray}
&&\mathcal{L}_{\bar{\nu}\chi^-\tilde{L}}=\bar{\nu}_i\Big((-g_2U^*_{j1}\sum_{a=1}^3U^{V*}_{ia}Z^E_{ka}+U^*_{j2}\sum_{a=1}^3U^{V*}_{ia}Y^a_lZ^E_{k(3+a)})P_L
\nonumber\\&&\hspace{1.6cm}+\sum_{a,b=1}^3Y_{\nu}^{ab}U^V_{i(3+a)}Z^E_{kb}V_{j2}P_R\Big)\chi^-_j\tilde{L}_k,
\\&&\mathcal{L}_{\bar{\chi}^0l\tilde{L}}=\bar{\chi}^0_i\Big\{\Big(\frac{1}{\sqrt{2}}(g_1N^*_{i1}+g_2N^*_{i2}+g_{YX}N^*_{i5})Z^E_{kj}
-N^*_{i3}Y^j_lZ^E_{k(3+j)}\Big)P_L\nonumber\\&&\hspace{1.6cm}
-\Big[\frac{1}{\sqrt{2}}\Big(2g_1N_{i1}+(2g_{YX}+g_X)N_{i5}\Big)Z^E_{k(3+j)}+Y_{l}^jZ^E_{kj}N_{i3}\Big]P_R\Big\}l_j\tilde{L}_k.
\end{eqnarray}

4. The $W$-related vertices.
\begin{eqnarray}
&&\mathcal{L}_{\tilde{L}\tilde{\nu}^{R*}W}=-\frac{1}{2}g_2\tilde{L}_i\tilde{\nu}^{R*}_j
\sum_{a=1}^3Z^{E*}_{ia}Z^{R*}_{ja}(-p_{\mu}^{\tilde{\nu}_j^R}+p_\mu^{\tilde{L}_i})W^\mu,
\\&&\mathcal{L}_{\tilde{L}\tilde{\nu}^{I*}W}=\frac{i}{2}g_2\tilde{L}_i\tilde{\nu}^{I*}_j
\sum_{a=1}^3Z^{E*}_{ia}Z^{I*}_{ja}(-p_{\mu}^{\tilde{\nu}_j^I}+p_\mu^{\tilde{L}_i})W^\mu,
\\&&\mathcal{L}_{\chi^0_i\chi^-_jW}=-\frac{i}{2}g_2\gamma_\mu\bar{\chi}^0_i{\chi}_j^- \big[(2U^*_{j1}N_{i2}+\sqrt{2}U^*_{j2}N_{i3})P_L+(2N^*_{i2}V_{j1}-\sqrt{2}N^*_{i4}V_{j2})P_R\big].
\end{eqnarray}

5. The quark-related vertices.
\begin{eqnarray}
&&\mathcal{L}_{\chi^0d\tilde{D}}=-\frac{i}{6}\bar{\chi}^0_i\Big\{\Big[\sqrt{2}(g_1 N_{1i}-3 g_2 N_{2i} + g_{Y X} N_{5i})Z^{\tilde{D}*}_{jk} +6 N_{3i} Y_d^j Z^{\tilde{D}*}_{(3+j)k} \Big]P_L\nonumber\\
&&\hspace{1.6cm}+\Big[6 Y_d^j Z^{\tilde{D}*}_{jk}  N^*_{3i}
+ \sqrt{2} Z^{\tilde{D}*}_{(3+j)k} [2 g_1 N^*_{1i} + (2 g_{YX} + 3 g_{X})N^*_{5i}]\Big]P_R\Big\}d_j\tilde{D}^*_k,
\end{eqnarray}
\begin{eqnarray}
&&\mathcal{L}_{\chi^0u\tilde{U}}=-\frac{i}{6}\bar{\chi}^0_i\Big\{ \Big[\sqrt{2}( g_1 N_{1i} +3  g_2 N_{2i} + g_{YX} N_{5i}) Z^{\tilde{U}*}_{jk}+6 N_{4i} Y_u^j Z^{\tilde{U}*}_{(3+j)k} \Big]P_L\nonumber\\
&&\hspace{1.6cm}- \Big[ \sqrt{2}Z^{\tilde{U}*}_{(3+j)k}  \Big((3 g_{X} + 4g_{Y X})N^*_{5i} + 4 g_1 N^*_{1i}\Big) -6 Y_u^j Z^{\tilde{U}*}_{jk} N^*_{4i}\Big] P_R\Big\}u_j\tilde{U}^*_k,
\\&&\mathcal{L}_{\chi^-d\tilde{U}}=\bar{d}_i\Big\{U^*_{j2}\sum_{a=1}^3 Z^{\tilde{U}*}_{ka} Y_d^a P_L +\Big[\sum_{a=1}^3 Y_u^a Z^{\tilde{U}*}_{k(3+a)}V_{j2}-g_2\sum_{a=1}^3 Z^{\tilde{U}*}_{ka}V_{j1}\Big]P_R\Big\}{\chi}^-_i\tilde{U}^*_k,
\\&&\mathcal{L}_{\chi^-u\tilde{D}}=\bar{\chi}^-_i\Big\{\Big[U^*_{i2} \sum_{a=1}^3 Y_d^a Z^{\tilde{D}}_{k(3+a)}-g_2 U^*_{i1} \sum_{a=1}^3 Z^{\tilde{D}}_{ka}\Big] P_L+\sum_{a=1}^3 Z^{\tilde{D}}_{ka} Y_u^{a*} V_{i2} P_R\Big\}{u}_j\tilde{D}^*_k.
\end{eqnarray}

\section{Formulation}
\subsection{Observables}
The effective Lagrangian of the $b\rightarrow c\ell\bar{\nu}^{_\ell} ~(\ell=e,~\mu,~\tau) $ process is
\begin{eqnarray}
&&\mathcal{L}_{eff}^{b\rightarrow c\ell\bar{\nu}^{_\ell}}=\sqrt{2}G_{F}V_{cb}(\mathcal{C}_{VL}^{\ell}\mathcal{O}_{VL}^{\ell}+\mathcal{C}_{VR}^{\ell}\mathcal{O}_{VR}^{\ell}+\mathcal{C}_{AL}^{\ell}\mathcal{O}_{AL}^{\ell}+\mathcal{C}_{AR}^{\ell}\mathcal{O}_{AR}^{\ell}+\mathcal{C}_{SL}^{\ell}\mathcal{O}_{SL}^{\ell}\nonumber\\&&\hspace{1.7cm}
+\mathcal{C}_{SR}^{\ell}\mathcal{O}_{SR}^{\ell}+\mathcal{C}_{PL}^{\ell}\mathcal{O}_{PL}^{\ell}+\mathcal{C}_{PR}^{\ell}\mathcal{O}_{PR}^{\ell}+\mathcal{C}_{TL}^{\ell}\mathcal{O}_{TL}^{\ell}+\mathcal{C}_{TR}^{\ell}\mathcal{O}_{TR}^{\ell}),
\label{EL}
\end{eqnarray}
and the full set of operators is~\cite{G}
\begin{eqnarray}
&&\mathcal{O}_{VL}^{\ell}=[\bar{c}\gamma^\mu b][\bar{\ell}\gamma_\mu P_L\nu^{_\ell}], ~~~~~~~~~~~~
\mathcal{O}_{VR}^{\ell}=[\bar{c}\gamma^\mu b][\bar{\ell}\gamma_\mu P_R\nu^{_\ell}],\nonumber\\&&\hspace{0cm}
\mathcal{O}_{AL}^{\ell}=[\bar{c}\gamma^\mu \gamma_5 b][\bar{\ell}\gamma_\mu P_L\nu^{_\ell}], ~~~~~~~~~~
\mathcal{O}_{AR}^{\ell}=[\bar{c}\gamma^\mu \gamma_5 b][\bar{\ell}\gamma_\mu P_R\nu^{_\ell}],\nonumber\\&&\hspace{0cm}
\mathcal{O}_{SL}^{\ell}=[\bar{c} b][\bar{\ell} P_L\nu^{_\ell}], ~~~~~~~~~~~~~~~~~~~
\mathcal{O}_{SR}^{\ell}=[\bar{c} b][\bar{\ell} P_R\nu^{_\ell}],\nonumber\\&&\hspace{0cm}
\mathcal{O}_{PL}^{\ell}=[\bar{c}\gamma_5 b][\bar{\ell} P_L\nu^{_\ell}], ~~~~~~~~~~~~~~~~
\mathcal{O}_{PR}^{\ell}=[\bar{c}\gamma_5 b][\bar{\ell} P_R\nu^{_\ell}],\nonumber\\&&\hspace{0cm}
\mathcal{O}_{TL}^{\ell}=[\bar{c}\sigma^{\mu\nu} b][\bar{\ell}\sigma_{\mu\nu} P_L\nu^{_\ell}], ~~~~~~~~~~
\mathcal{O}_{TR}^{\ell}=[\bar{c}\sigma^{\mu\nu} b][\bar{\ell}\sigma_{\mu\nu} P_R\nu^{_\ell}].
\label{SF}
\end{eqnarray}
Combined with the operators in Eq.(\ref{SF}), we can derive the Wilson coefficients of the effective Lagrangian from the amplitude. Then Wilson coefficients are used to calculate the observable.

The observable $R_{D^{(*)}}$ is defined as
\begin{eqnarray}
&&R_{D^{(*)}}=\frac{\mathcal{B}_{\tau}^{D^{(*)}}}{\mathcal{B}_{l}^{D^{(*)}}}=\frac{\mathcal{B}(\overline{B}\rightarrow D^{(*)}\tau \bar{\nu}_{\tau})}{\mathcal{B}(\overline{B}\rightarrow D^{(*)}l \bar{\nu}_{l})}.
\label{DFRDX}
\end{eqnarray}
The formula of the process $\overline{B}\rightarrow D^{(*)}l \bar{\nu}_{\ell}$ is given by~\cite{A}
\begin{eqnarray}
&&\mathcal{B}_{\ell}^{D^{(*)}}=\int \mathcal{N}|p_{D^{(*)}}|(2a_{\ell}^{D^{(*)}}+\frac{2}{3}c_{\ell}^{D^{(*)}})dq^{2}.
\label{BB}
\end{eqnarray}
 The interval of this integral is $[m_{\ell}^{2},(M_{B}-M_{D^{(*)}})^{2}]$. In Eq.(\ref{BB}), $q^{2}$ is the invariant mass squared of the lepton-neutrino system, and the normalisation factor $\mathcal{N}$ is represented by the following formula:
\begin{eqnarray}
&&\mathcal{N}=\frac{\tau_{B}G_{F}^{2}|V_{cb}|^{2}q^{2}}{256\pi^{3}M_{B}^{2}}(1-\frac{m_{\ell}^{2}}{q^{2}})^2.
\end{eqnarray}
Here $\tau_{B}$ is the lifetime of the $B$ meson, $G_{F}$ is the Fermi coupling constant, and $|p_{D^{(*)}}|$ is the absolute value of the $D^{(*)}$ meson momentum. $|p_{D^{(*)}}|$ is given by
\begin{eqnarray}
&&|p_{D^{(*)}}|=\frac{\sqrt{(M_{B}^{2})^{2}+(M_{D^{(*)}}^{2})^{2}+(q^{2})^{2}-2(M_{B}^{2}M_{D^{(*)}}^{2}+M_{D^{(*)}}^{2}q^{2}+q^{2}M_{B}^{2})}}{2M_{B}}.
\end{eqnarray}

The expressions for $a_{\ell}^{D}$ and $c_{\ell}^{D}$ are~\cite{A}
\begin{eqnarray}
&&a_{\ell}^{D}=8\Big\{\frac{M_{B}^{2}|p_{D}|^{2}}{q^{2}}\big(|\mathcal{C}_{VL}^{\ell}|^{2}+|\mathcal{C}_{VR}^{\ell}|^{2}\big)\textbf{F}_{+}^{2}+\frac{(M_{B}^{2}- M_{D}^{2})^{2}}{4(m_{b}-m_{c})^{2}}\big(|\mathcal{C}_{SL}^{\ell}|^{2}+|\mathcal{C}_{SR}^{\ell}|^{2}\big)\textbf{F}_{0}^{2}\nonumber\\&&\hspace{1cm}
+m_{\ell}\Big[\frac{(M_{B}^{2}- M_{D}^{2})^{2}}{2q^{2}(m_{b}-m_{c})}\big(\mathcal{R}(\mathcal{C}_{SL}^{\ell}\mathcal{C}_{VL}^{\ell*})+\mathcal{R}(\mathcal{C}_{SR}^{\ell}\mathcal{C}_{VR}^{\ell*})\big)\textbf{F}_{0}^{2}\nonumber\\&&\hspace{1cm}
+\frac{4M_{B}^{2} |p_{D}|^{2} }{q^{2}(M_{B}+M_{D})}\big(\mathcal{R}(\mathcal{C}_{TL}^{\ell}\mathcal{C}_{VL}^{\ell*})+\mathcal{R}(\mathcal{C}_{TR}^{\ell}\mathcal{C}_{VR}^{\ell*})\big)\textbf{F}_{+}\textbf{F}_{T}\Big]\nonumber\\&&\hspace{1cm}
+m_{\ell}^{2}\Big[\frac{(M_{B}^{2}- M_{D}^{2})^{2}}{4q^{4}}\big(|\mathcal{C}_{VL}^{\ell}|^{2}+|\mathcal{C}_{VR}^{\ell}|^{2}\big)\textbf{F}_{0}^{2}\nonumber\\&&\hspace{1cm}
+\frac{4|p_{D}|^{2}M_{B}^{2}  }{q^{2}(M_{B}+M_{D})^{2}}\big(| \mathcal{C}_{TL}^{\ell}|^{2}+| \mathcal{C}_{TR}^{\ell}|^{2}\big)\textbf{F}_{T}^{2}\Big]\Big\},
\\
&&c_{\ell}^{D}=8\Big\{\frac{4M_{B}^{2}|p_{D}|^{2}}{(M_{B}+ M_{D})^{2}}\big(|\mathcal{C}_{TL}^{\ell}|^{2}+|\mathcal{C}_{TR}^{\ell}|^{2}\big)\textbf{F}_{T}^{2}\nonumber\\&&\hspace{1cm}
-\frac{M_{B}^{2} |p_{D}|^{2}}{q^{2}}\big(|\mathcal{C}_{VL}^{\ell}|^{2}+|\mathcal{C}_{VR}^{\ell}|^{2}\big)\textbf{F}_{+}^{2}
+m_{\ell}^{2}\Big[\frac{|p_{D}|^{2}M_{B}^{2}}{q^{4}}\big(|\mathcal{C}_{VL}^{\ell}|^{2}+|\mathcal{C}_{VR}^{\ell}|^{2}\big)\textbf{F}_{+}^{2}\nonumber\\&&\hspace{1cm}
-\frac{4|p_{D}|^{2}M_{B}^{2}}{(M_{B}+M_{D})^{2}q^{2}}\big(|\mathcal{C}_{TL}^{\ell}|^{2}+| \mathcal{C}_{TR}^{\ell}|^{2}\big)\textbf{F}_{T}^{2}\Big]\Big\}.
\end{eqnarray}
The expressions of form factors $F_0(q^2)$ and $F_+(q^2)$ \cite{A} are given by lattice QCD techniques,
\begin{eqnarray}
F_+(z) = \frac{1}{\phi_+(z)} \sum_{k=0}^3 a_k^+ \, z^k ,\,~~~ F_0(z) = \frac{1}{\phi_0(z)} \sum_{k=0}^3 a_k^0 \, z^k \, ,
\end{eqnarray}
where
\begin{eqnarray}
z \equiv  z(q^2)  = \frac{\sqrt{(M_B + M_D)^2 - q^2} - \sqrt{4 M_B M_D}}{\sqrt{(M_B + M_D)^2 - q^2} + \sqrt{4 M_B M_D}} \, .
\end{eqnarray}
The functions $\phi_+(z)$ and $\phi_0(z)$ are as follows,
\begin{eqnarray}
\phi_+(z) &=& 1.1213  \frac{(1+z)^2 (1-z)^{1/2}}{\left[(1+r)(1-z) + 2 \sqrt{r}(1+z)\right]^5} \, , \\
\phi_0(z) &=& 0.5299 \frac{(1+z)(1 - z)^{3/2}}{\left[(1+r)(1-z) + 2 \sqrt{r} (1+z)\right]^4} \, ,
\end{eqnarray}
where $r = M_D/M_B$. The form factor $F_T$ is not calculated by lattice QCD techniques. But according to the Ref. \cite{fromf}, we can give the expression for $F_T$ as follows,
\begin{eqnarray}
F_T(q^2) &=& \frac{0.69}{\left(1- \frac{q^2}{(6.4 {\rm GeV})^2} \right) \left(1 - 0.56 \frac{q^2}{(6.4 {\rm GeV})^2}\right)}\,.
\end{eqnarray}

 $a_{\ell}^{D^{*}}$, $c_{\ell}^{D^{*}}$ and the related form factors can be found in Refs. \cite{A,36}. To save space in the text, their specific forms are shown in Appendix \ref{alcl}.

\subsection{Feynman diagrams}

\begin{figure}[ht]
\centering
\subfigure[]{
\setlength{\unitlength}{5.0mm}
\includegraphics[width=1.5in]{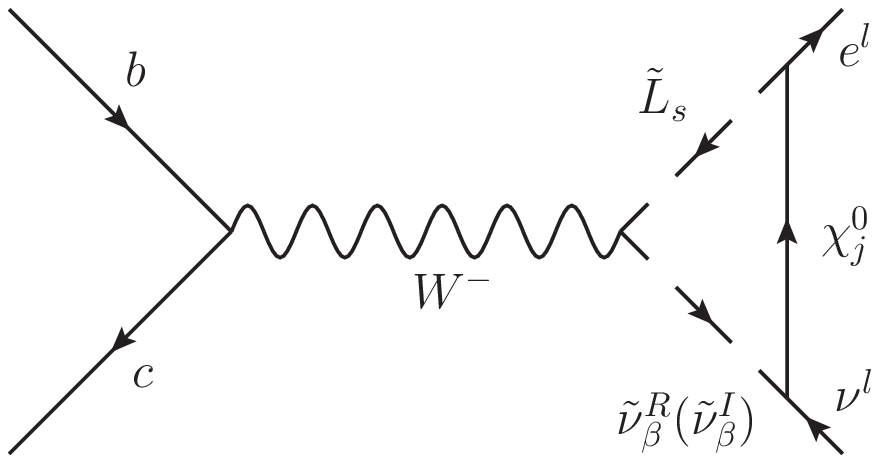}
\label{Fig1}
}
\subfigure[]{
\setlength{\unitlength}{5.0mm}
\includegraphics[width=1.5in]{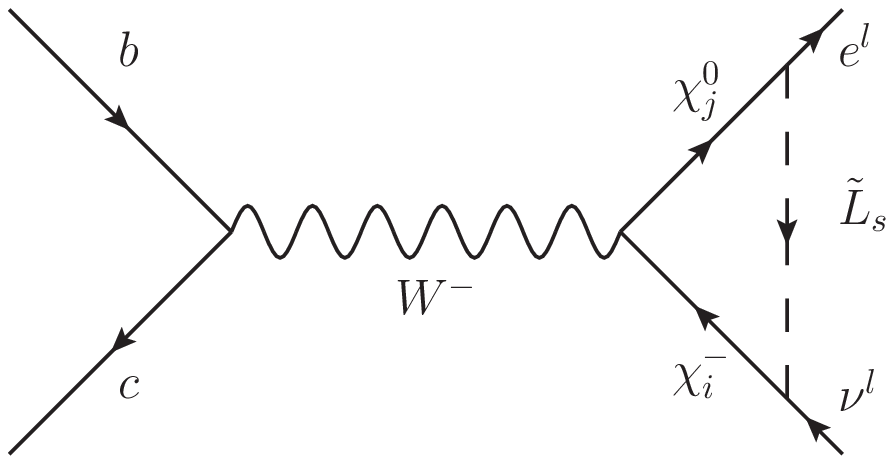}
\label{Fig2}
}
\subfigure[]{
\setlength{\unitlength}{5.0mm}
\includegraphics[width=1.5in]{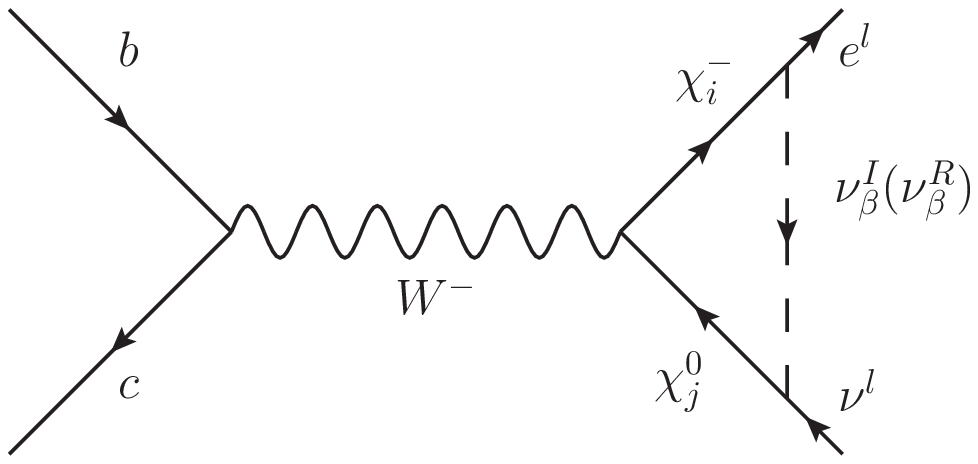}
\label{Fig3}
}
\caption{The penguin-type Feynman diagrams that can correct $R_{D^{(*)}}$.}\label{N1}
\end{figure}
\begin{figure}[ht]
\centering
\subfigure[]{
\setlength{\unitlength}{5.0mm}
\includegraphics[width=1.4in]{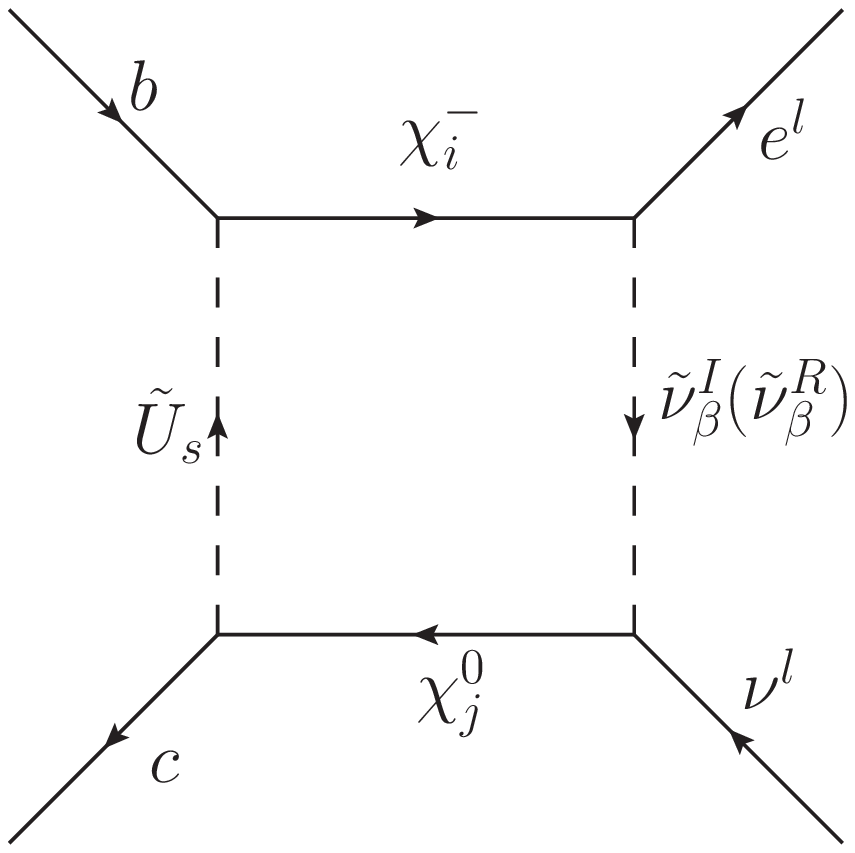}
\label{Fig4}
}
\subfigure[]{
\setlength{\unitlength}{5.0mm}
\includegraphics[width=1.4in]{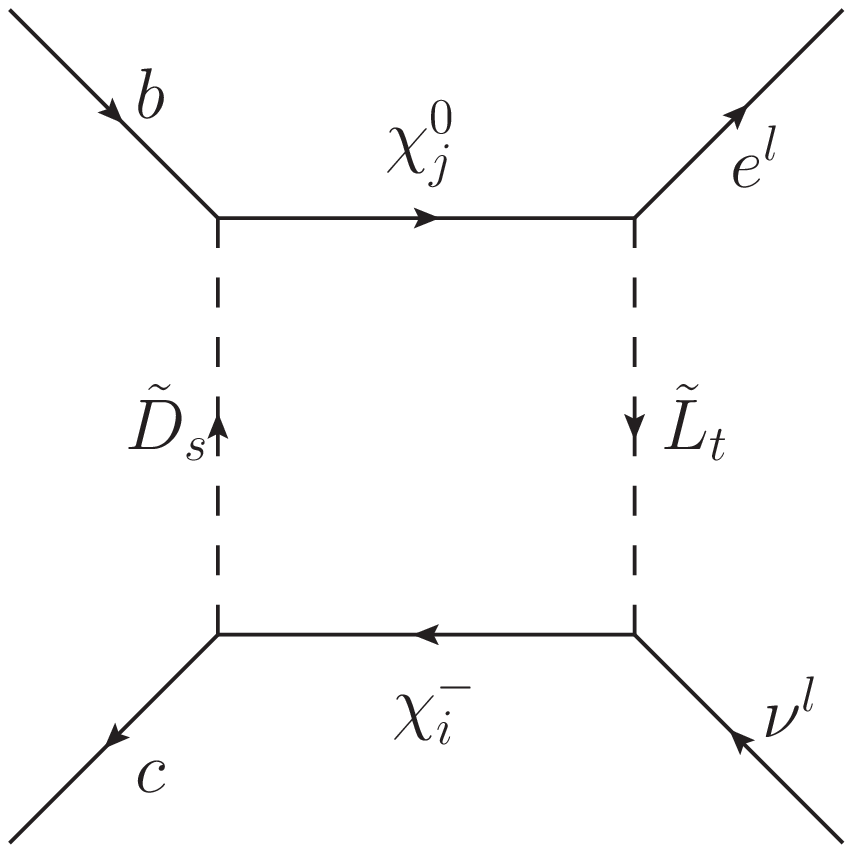}
\label{Fig5}
}
\subfigure[]{
\setlength{\unitlength}{5.0mm}
\includegraphics[width=1.4in]{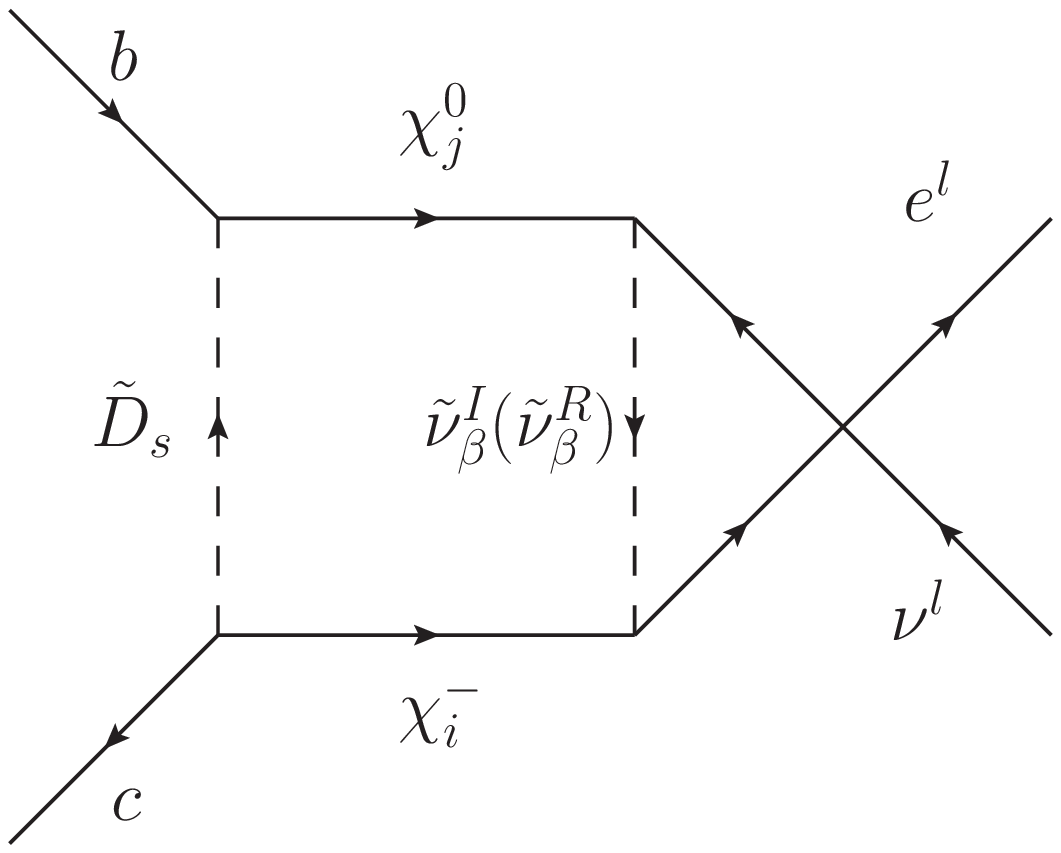}
\label{Fig6}
}
\subfigure[]{
\setlength{\unitlength}{5.0mm}
\includegraphics[width=1.4in]{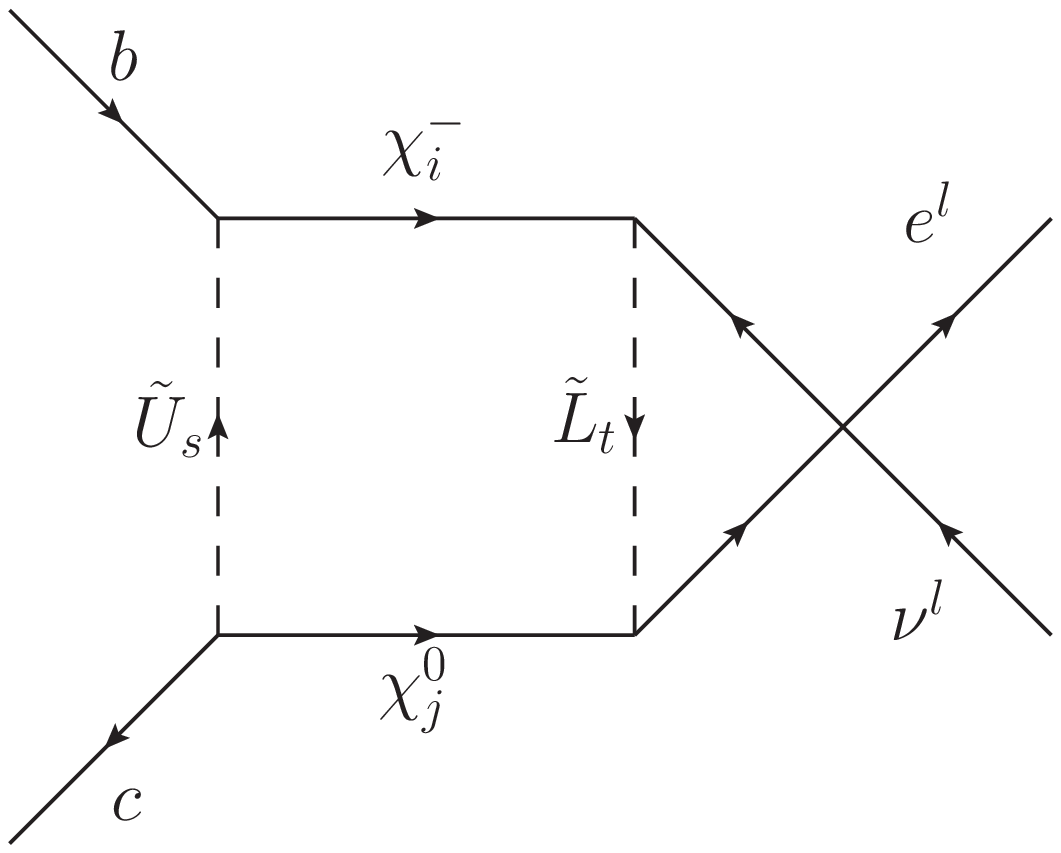}
\label{Fig7}
}
\caption{The box-type Feynman diagrams that can correct $R_{D^{(*)}}$.}\label{N2}
\end{figure}

Fig.~\ref{N1} and Fig.~\ref{N2} show the one-loop Feynman diagrams under the $U(1)_X$SSM.
The Feynman diagrams in Fig.~\ref{N1} are all UV divergent.  Through $\overline{MS}$ subtraction, we can remove the infinite
terms of the Wilson coefficients obtained from the amplitude. Therefore, taking Fig.~\ref{Fig1} as an example, we select the case where CP-odd sneutrinos exist and obtain the specific expression for the amplitude as follows:
\begin{eqnarray}
\mathcal{C}_{VL(1a)}^{\ell}&=&-[\sum_{\beta,s=1}^{6}\sum_{j=1}^{8}\frac{\mathcal{B}_{1}^{\ell sj}\mathcal{A}_{2}^{\beta\ell j}\mathcal{A}_{3}^{\beta s}\mathcal{A}_{4}}{m_{W}^{2}}\frac{1}{64\pi^{2}}F_{12}(x_{\chi^0_j},x_{\tilde{L}_{s}},x_{\tilde{\nu}^{I}_{\beta}} )]/(\sqrt{2}G_{F}V_{cb}),\nonumber\\&&\hspace{-2cm}
\mathcal{C}_{AL(1a)}^{\ell}~=-\mathcal{C}_{VL(a)}^{\ell}.
\label{Ca}
\end{eqnarray}

The specific expressions of couplings, $F_{12}$ and $F_{11}$(Although the above formula does not use $F_{11}$, it is needed to calculate the other two diagrams.) are as follows:
\begin{eqnarray}
&&\mathcal{A}_{3}^{\beta s}=-\frac{1}{2}g_2\sum_{a=1}^3Z^{E*}_{ka}Z^{R*}_{\beta a},\hspace{1cm}\mathcal{A}_{4}=-\frac{1}{\sqrt{2}}g_2V_{cb},\nonumber\\&&\hspace{0cm}
\mathcal{B}_{1}^{\ell sj}=\frac{1}{\sqrt{2}}(g_1N_{j1}+g_2N_{j2}+g_{YX}N_{j5})Z^{E*}_{s \ell}
-N_{j3}Y^\ell_lZ^{E*}_{s(3+\ell)},\nonumber\\&&\hspace{0cm}
\mathcal{A}_{2}^{\beta\ell j}=(g_2N^*_{j2}-g_{YX}N^*_{j5}-g_1N^*_{j1}) \sum_{a=1}^3Z^{I*}_{\beta a}U_{\ell a}^{V*},\nonumber\\&&\hspace{0cm}\nonumber
\end{eqnarray}
\begin{eqnarray}
&&F_{11}(x_{1},x_{2},x_{3})=\frac{x_{1}\ln x_{1}}{(x_{1}-x_{2})(x_{1}-x_{3})}+\frac{x_{2}\ln x_{2}}{(x_{2}-x_{1})(x_{2}-x_{3})}+\frac{x_{3}\ln x_{3}}{(x_{3}-x_{1})(x_{3}-x_{2})},
\nonumber\\&&\hspace{0cm}
F_{12}(x_{1},x_{2},x_{3})=\frac{x_{1}^2\ln x_{1}}{(x_{1}-x_{2})(x_{1}-x_{3})}+\frac{x_{2}^2\ln x_{2}}{(x_{2}-x_{1})(x_{2}-x_{3})}+\frac{x_{3}^2\ln x_{3}}{(x_{3}-x_{1})(x_{3}-x_{2})}.
\end{eqnarray}
According to the above procedure, we can get the coefficients of all the diagrams in Fig.~\ref{N1}.

Compared to the penguin-type Feynman diagrams in Fig.~\ref{N1}, the box-type Feynman diagrams are convergent. Here again, the case of CP-odd sneutrino in Fig.~\ref{Fig4} is taken as an example, and its amplitude is expressed in the following form:
\begin{eqnarray}
\mathcal{C}_{VL(2a)}^{\ell}&=&\sum_{\beta,s=1}^{6}\sum_{i=1}^{2}\sum_{j=1}^{8}
\big[\frac{\mathcal{B}_{5}^{s i}\mathcal{B}_{6}^{\beta\ell i}\mathcal{A}_{7}^{s j}\mathcal{A}_{1}^{\beta\ell j}m_{\chi^0_{j}}m_{\chi^-_{i}}}{m_{W}^{4}}\frac{1}{64\pi^{2}}
F_{21}(x_{\tilde{\nu}^{I}_{\beta}},x_{\chi^0_{j}},x_{\tilde{U}_{s}},x_{\chi^-_{i}})\nonumber\\&&
-\frac{\mathcal{A}_{5}^{s i}\mathcal{B}_{6}^{\beta\ell i}\mathcal{B}_{7}^{s j}\mathcal{A}_{1}^{\beta\ell j}}{m_{W}^{2}}\frac{1}{128\pi^{2}}F_{22}(x_{\tilde{\nu}^{I}_{\beta}},
x_{\chi^0_{j}},x_{\tilde{U}_{s}},x_{\chi^-_{i}})\big]/(\sqrt{2}G_{F}V_{cb}),
\label{CVLe}
\\
\mathcal{C}_{AL(2a)}^{\ell}&=&\sum_{\beta,s=1}^{6}\sum_{i=1}^{2}\sum_{j=1}^{8}\big[\frac{\mathcal{B}_{5}^{s i}\mathcal{B}_{6}^{\beta\ell i}\mathcal{A}_{7}^{s j}\mathcal{A}_{1}^{\beta\ell j}m_{\chi^0_{j}}m_{\chi^-_{i}}}{m_{W}^{4}}\frac{1}{64\pi^{2}}F_{21}(x_{\tilde{\nu}^{I}_{\beta}},
x_{\chi^0_{j}},x_{\tilde{U}_{s}},x_{\chi^-_{i}})\nonumber\\&&\hspace{0cm}
+\frac{\mathcal{A}_{5}^{s i}\mathcal{B}_{6}^{\beta\ell i}\mathcal{B}_{7}^{s j}\mathcal{A}_{1}^{\beta\ell j}}{m_{W}^{2}}\frac{1}{128\pi^{2}}F_{22}
(x_{\tilde{\nu}^{I}_{\beta}},
x_{\chi^0_{j}},x_{\tilde{U}_{s}},x_{\chi^-_{i}})\big]/(\sqrt{2}G_{F}V_{cb}),
\label{CALe}
\\
\mathcal{C}_{SL(2a)}^{\ell}&=&\sum_{\beta,s=1}^{6}\sum_{i=1}^{2}\sum_{j=1}^{8}\big[\frac{\mathcal{A}_{5}^{s i}\mathcal{A}_{6}^{\beta\ell i} \mathcal{A}_{7}^{s j}\mathcal{A}_{1}^{\beta\ell j}m_{\chi^0_{j}}m_{\chi^-_{i}}}{m_{W}^{4}}\frac{1}{64\pi^{2}}F_{21}(x_{\tilde{\nu}^{I}_{\beta}},
x_{\chi^0_{j}},x_{\tilde{U}_{s}},x_{\chi^-_{i}})\nonumber\\&&\hspace{0cm}
+\frac{\mathcal{B}_{5}^{s i}\mathcal{A}_{6}^{\beta\ell i}\mathcal{B}_{7}^{s j}\mathcal{A}_{1}^{\beta\ell j}}{m_{W}^{2}}
\frac{1}{64\pi^{2}}F_{22}(x_{\tilde{\nu}^{I}_{\beta}},
x_{\chi^0_{j}},x_{\tilde{U}_{s}},x_{\chi^-_{i}})\big]/(\sqrt{2}G_{F}V_{cb}),
\label{CSLe}
\\
\mathcal{C}_{TL(2a)}^{\ell}&=&\sum_{\beta,s=1}^{6}\sum_{i=1}^{2}\sum_{j=1}^{8}
\frac{\mathcal{A}_{5}^{s i}\mathcal{A}_{6}^{\beta\ell i}\mathcal{A}_{7}^{s j}\mathcal{A}_{1}^{\beta\ell j}m_{\chi^0_{j}}m_{\chi^-_{i}}}{128\sqrt{2}\pi^{2}m_{W}^{4}G_{F}V_{cb}}F_{21}(x_{\tilde{\nu}^{I}_{\beta}},
x_{\chi^0_{j}},x_{\tilde{U}_{s}},x_{\chi^-_{i}}),
\label{CTLe}
\end{eqnarray}
\begin{eqnarray}
\mathcal{C}_{PL(2a)}^{\ell}&=&\sum_{\beta,s=1}^{6}\sum_{i=1}^{2}\sum_{j=1}^{8}\big[\frac{(\mathcal{A}_{5}^{s i}\mathcal{A}_{6}^{\beta\ell i}-\mathcal{B}_{5}^{s i}\mathcal{B}_{6}^{\beta\ell i})\mathcal{A}_{7}^{s j}\mathcal{A}_{1}^{\beta\ell j}m_{\chi^0_{j}}m_{\chi^-_{i}}}{m_{W}^{4}{128\pi^{2}}}F_{21}(x_{\tilde{\nu}^{I}_{\beta}},
x_{\chi^0_{j}},x_{\tilde{U}_{s}},x_{\chi^-_{i}})\nonumber\\&&\hspace{0cm}
+\frac{\mathcal{B}_{5}^{s i}\mathcal{A}_{6}^{\beta\ell i}\mathcal{B}_{7}^{s j}\mathcal{A}_{1}^{\beta\ell j}}{m_{W}^{2}}\frac{1}{64\pi^{2}} F_{22}(x_{\tilde{\nu}^{I}_{\beta}},
x_{\chi^0_{j}},x_{\tilde{U}_{s}},x_{\chi^-_{i}})\big]/(\sqrt{2}G_{F}V_{cb}),
\label{CPLe}
\\
\mathcal{C}_{PR(2a)}^{\ell}&=&-\sum_{\beta,s=1}^{6}\sum_{i=1}^{2}\sum_{j=1}^{8}\frac{(\mathcal{A}_{5}^{s i}\mathcal{A}_{6}^{\beta\ell i}+\mathcal{B}_{5}^{s i}\mathcal{B}_{6}^{\beta\ell i})\mathcal{A}_{7}^{s j}\mathcal{A}_{1}^{\beta\ell j}m_{\chi^0_{j}}m_{\chi^-_{i}}}{128\sqrt{2}\pi^{2}m_{W}^{4}G_{F}V_{cb}}F_{21}(x_{\tilde{\nu}^{I}_{\beta}},
x_{\chi^0_{j}},x_{\tilde{U}_{s}},x_{\chi^-_{i}}).
\label{CPRe}
\end{eqnarray}

The specific expressions of $F_{21}$, $F_{22}$ and couplings are as follows:
\begin{eqnarray}
&&\mathcal{A}_{5}^{s i}=\sum_{a=1}^3 Y_u^a Z^{\tilde{U}*}_{k(3+a)}V_{j2}-g_2\sum_{a=1}^3 Z^{\tilde{U}*}_{ka}V_{j1},\hspace{1.3cm}\mathcal{B}_{5}^{s i}=U^*_{j2}\sum_{a=1}^3 Z^{\tilde{U}*}_{ka} Y_d^a ,\nonumber\\
&&\mathcal{A}_{6}^{\beta\ell i}=\frac{1}{\sqrt{2}}U^*_{j2}Z^{I*}_{ki}Y_l^i,\hspace{3.3cm}
\mathcal{B}_{6}^{\beta\ell i}=-\frac{1}{\sqrt{2}}g_2V_{j1}Z^{I*}_{ki},\nonumber\\
&&\mathcal{A}_{7}^{s j}=\frac{\sqrt{2}}{6}Z^{\tilde{U}}_{(3+j)k} (3 g_{X}N_{5i} + 4g_{Y X}N_{5i} + 4 g_1 N_{1i}) -6 Y_u^j Z^{\tilde{U}}_{jk} N_{4i},\nonumber\\&&
\mathcal{B}_{7}^{s j}=-\frac{\sqrt{2}}{6}( g_1 N^*_{1i} +3  g_2 N^*_{2i} + g_{YX} N^*_{5i}) Z^{\tilde{U}}_{jk}+6 N^*_{4i} Y_u^j Z^{\tilde{U}}_{(3+j)k} ,\nonumber\\&&
\mathcal{A}_{1}^{\beta\ell j}=(-g_2N^*_{i2}+g_{YX}N^*_{i5}+g_1N^*_{i1})\sum_{a=1}^3Z^{I*}_{ka}U_{ja}^{V*},\nonumber\\&&
F_{21}(x_{1},x_{2},x_{3},x_{4})=\frac{x_{1}\ln x_{1}}{(x_{1}-x_{2})(x_{1}-x_{3})(x_{1}-x_{4})}+\frac{x_{2}\ln x_{2}}{(x_{2}-x_{1})(x_{2}-x_{3})(x_{2}-x_{4})}\nonumber\\&&\hspace{3.3cm}
+\frac{x_{3}\ln x_{3}}{(x_{3}-x_{1})(x_{3}-x_{2})(x_{3}-x_{4})}+\frac{x_{4}\ln x_{4}}{(x_{4}-x_{1})(x_{4}-x_{2})(x_{4}-x_{3})},\nonumber\\&&\hspace{0cm}
F_{22}(x_{1},x_{2},x_{3},x_{4})=\frac{x_{1}^{2}\ln x_{1}}{(x_{1}-x_{2})(x_{1}-x_{3})(x_{1}-x_{4})}+\frac{x_{2}^{2}\ln x_{2}}{(x_{2}-x_{1})(x_{2}-x_{3})(x_{2}-x_{4})}\nonumber\\&&\hspace{3.3cm}
+\frac{x_{3}^{2}\ln x_{3}}{(x_{3}-x_{1})(x_{3}-x_{2})(x_{3}-x_{4})}+\frac{x_{4}^{2}\ln x_{4}}{(x_{4}-x_{1})(x_{4}-x_{2})(x_{4}-x_{3})}.
\end{eqnarray}

By combining all the results of the penguin-type Feynman diagrams and the box-type Feynman diagrams, we can get all the non-zero Wilson coefficients, and then get observation $R_{D^{(*)}}$.

\section{numerical results}
In this section of the numerical results, we consider multiple experimental constraints including:

   1. The lightest CP-even Higgs $h^0$ mass is around 125.25 GeV \cite{pdg}.

   2. The $Z^\prime$ boson mass is larger than 5.1 TeV to satisfy LHC experiments \cite{mzp}.

   3. The ratio between $M_{Z^\prime}$ and its gauge $M_{Z^\prime}/g_X \geq 6 ~{\rm TeV}$ \cite{mzpbgx}.

   4. The new angle $\beta_\eta$ is constrained by LHC as $\tan \beta_\eta< 1.5$ \cite{tbn15}.

   5. The limitations for the particle masses accord to the PDG~\cite{pdg} data,
   and the concrete contents are the following. The neutralino mass is limited to more than $116~\rm{GeV}$,
   and the chargino mass is limited to more than $1000~\rm{GeV}$.
   The slepton mass is limited to more than $600~\rm{GeV}$.
   On the other hand, the squark mass is maintained at the $\rm{TeV}$ order of magnitude.

   6. The Higgs $h^0$ decays($h^0\rightarrow \gamma+\gamma,~ Z+Z,~ W+W,~ b+\bar{b},~\tau+\bar{\tau}$) \cite{pdg} should be satisfied.

   7. Muon anomalous magnetic dipole moment is also taken into account\cite{UX3,BNFN}.

We use images to visualize the effects of variables on the results. Under the premise of meeting the above experimental limitations, we use the parameters as follows:
\begin{eqnarray}
&&\tan\beta_{\eta}=0.9,~~\lambda_C=-0.2,~~\lambda_H=0.1,~~\mu=1.1~{\rm TeV},~~
v_S=4~{\rm TeV},
\nonumber\\&& M_{BB'}=0.35~{\rm TeV},~~ M_{BL}=1.8~{\rm TeV},~~M_S=({T_e})_{ii}=1.2~{\rm TeV},~~\kappa=0.05,
\nonumber\\&&({M_\nu}^2)_{11}=3.8~{\rm TeV}^2,~~({M_\nu}^2)_{22}=2.6~{\rm TeV}^2,~~({M_\nu}^2)_{33}=0.54~{\rm TeV}^2,\nonumber\\&&
({M_e}^2)_{11}=({M_e}^2)_{22}=5~{\rm TeV}^2,~~({T_x})_{ii}=-1~{\rm TeV},~
({m_{\tilde{D}}^2})_{ii}=2.7^2~{\rm{TeV}^2},\nonumber\\&&
({m_{\tilde{Q}}^2})_{ii}=({m_{\tilde{U}}^2})_{ii}=1.9^2~{\rm{TeV}^2},
~~({T_\nu})_{ii}=({T_u})_{ii}=({T_d})_{ii}=1~{\rm TeV},
\end{eqnarray}
%\begin{eqnarray}
%&&\tan\beta_{\eta}=0.9,~~\lambda_C=-0.2,~~\lambda_H=0.1,~~\mu=1.1~{\rm TeV},~~
%v_S=4~{\rm TeV},
%\nonumber\\&&kappa=0.05,~~ M_{BB'}=0.35~{\rm TeV},~~ M_{BL}=1.8~{\rm TeV},~~({m_{\tilde{D}}^2})_{ii}=2.7^2~{\rm{TeV}^2},
%\nonumber\\&&M_S=({T_e})_{ii}=1.2~{\rm TeV},~~ ({M_e}^2)_{11}=({M_e}^2)_{22}=5~{\rm TeV}^2,~~\left(M^2_e\right)_{33} = 4~{\rm TeV}^2,
%\nonumber\\&&({M_\nu}^2)_{11}=3.8~{\rm TeV}^2,~~({M_\nu}^2)_{22}=2.6~{\rm TeV}^2,~~({M_\nu}^2)_{33}=0.54~{\rm TeV}^2,\nonumber\\&&
%({m_{\tilde{Q}}^2})_{ii}=({m_{\tilde{U}}^2})_{ii}=1.9^2~{\rm{TeV}^2},
%~~({T_\nu})_{ii}=({T_u})_{ii}=({T_d})_{ii}=1~{\rm TeV},
%\nonumber\\&& \left(M^2_l\right)_{11} = 0.3~{\rm TeV}^2,
%~~\left(M^2_l\right)_{22} = \left(M^2_l\right)_{33} = 0.15~{\rm TeV}^2,~~({T_x})_{ii}=-1~{\rm TeV}.
%\end{eqnarray}
in which, $i=1,2,3$. And the value of $Y_{\nu}$ is in the order of magnitude of $10^{-8}$ to $10^{-6}$, which has little effect on the results and does not been made too much discussion.

The central values and uncertainties for the parameters $a_k^0$ and $a_k^+$ are shown in
table~\ref{B2D-FF-numbers}.
The other parameters like $a_k^0$ and $a_k^+$ are not shown here, and can be found in Ref. \cite{A}.
\begin{table}[h!]
\centering
\begin{tabular}{|l|cccc|cccc|}
\hline
&$a_0^+$& $a_1^+$& $a_2^+$& $a_3^+$& $a_0^0$& $a_1^0$& $a_2^0$& $a_3^0$ \\
\hline
Values & 0.01261 & -0.0963 & 0.37& -0.05 & 0.01140 & -0.0590 & 0.19 & -0.03 \\
Uncertainties & 0.00010 & 0.0033 & 0.11 & 0.90 & 0.00009 & 0.0028 & 0.10 & 0.87 \\
\hline
\end{tabular}
\caption{The central values and uncertainties for the parameters $a_k^0$ and $a_k^+$.\label{B2D-FF-numbers}}
\label{as}
\end{table}

\subsection{The one-dimensional graph}

\begin{figure}[!htbp]
\begin{center}
\begin{minipage}[c]{0.48\textwidth}
\includegraphics[width=2.8in]{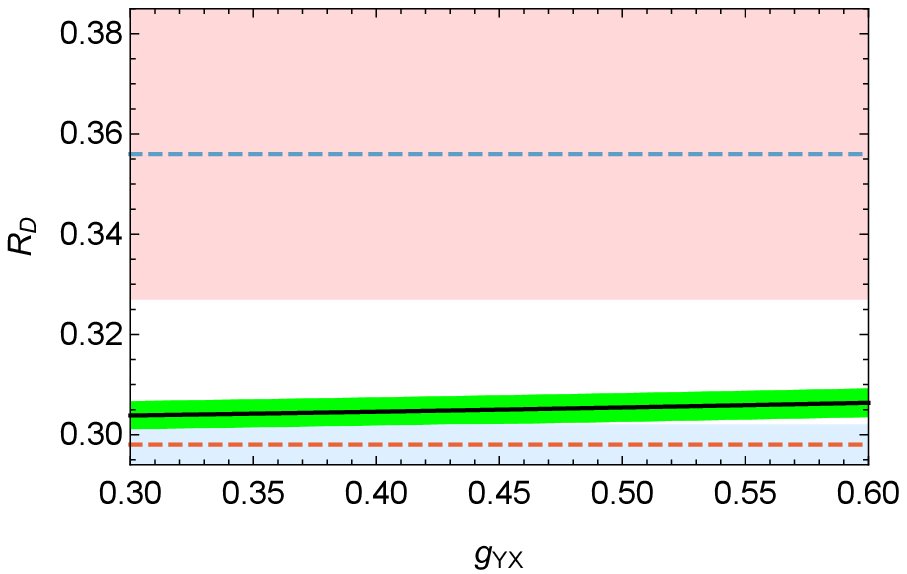}
\end{minipage}
\begin{minipage}[c]{0.48\textwidth}
\includegraphics[width=2.8in]{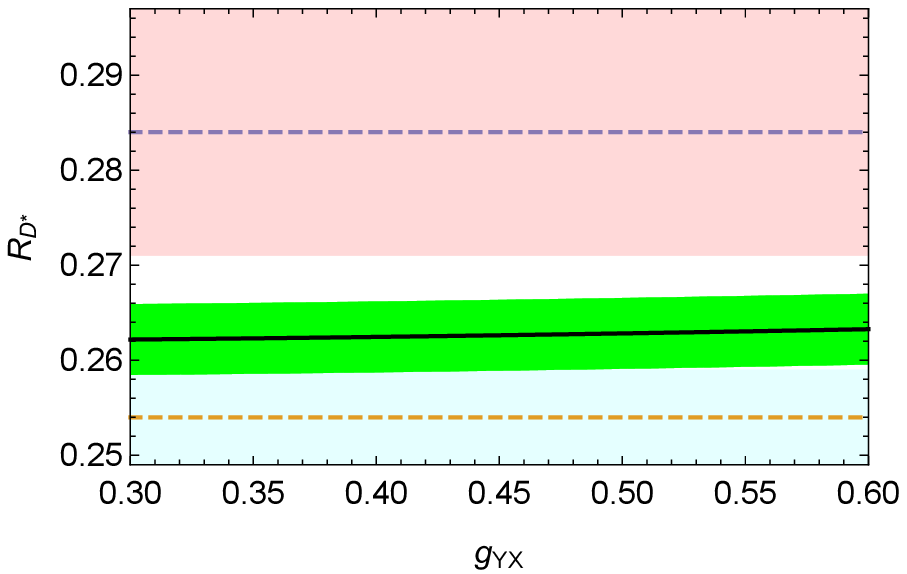}
\end{minipage}
\caption[]{\label{Fig8} The diagrams of $R_{D}$ (left) and $R_{D^{*}}$ (right) versus $g_{YX}$.}
\end{center}
\end{figure}

In this subsection, we use the parameters as $m_1=0.3~{\rm TeV},~m_2=1~{\rm TeV},~\tan\beta=57,~g_X=0.35,~({M_l}^2)_{11}=0.5~{\rm TeV}^2,~({M_l}^2)_{22}=0.05~{\rm TeV}^2,~({M_l}^2)_{33}=0.1~{\rm TeV}^2,~({M_e}^2)_{33}=4.5~{\rm TeV}^2$.
As a unique parameter in $U(1)_X$SSM, $g_{YX}$ is the coupling constant of gauge mixing that
affects the strength of coupling vertexes. To study its influence on objective measurement,
we vary $g_{YX}$ in the range of ($0.3-0.6$).
 By processing the data, we plot $g_{YX}$ versus $R_{D}$ (left) and $R_{D^{*}}$ (right) in Fig.~\ref{Fig8}.

The light cyan area and light red area represent one $\sigma$ range of the SM prediction and the experimental value. Among them, the orange dotted line and the blue dotted line represent the theoretical central value of SM and
the experimental central value. Our value of $R_{D^{(*)}}$ has exceeded the range of $\sigma_{SM}$, which can compensate for the deviation to a large extent. We can observe that the value of $R_{D^{(*)}}$ increases with the increase of $g_{YX}$. For the central value under $U(1)_X$SSM,
  it is expressed by a black line in the figure. When $g_{YX}$ reaches $0.6$, the value of $R_{D}$ reaches $0.306$ and the value of $R_{D^*}$ reaches $0.263$. When $g_{YX}$ is $0.3$, $R_{D}$ and $R_{D^*}$ are $0.303$ and $0.262$, respectively.
Considering the uncertainty of the form factor, we use a green area to represent the errors.
At this time, the maximum value of $R_{D}$ and $R_{D^*}$ can reach 0.309 and 0.267, respectively. Although it is still not in one $\sigma$ range of experimental values, it indeed improves the theoretical results. Also, it is obvious that the slope of the straight line in the left diagram is larger than that in the right diagram. Compared with the influence on $R_{D^{*}}$, the influence of $g_{YX}$ on $R_{D}$ is more obvious.

\begin{figure}[!htbp]
\begin{center}
\begin{minipage}[c]{0.48\textwidth}
\includegraphics[width=2.8in]{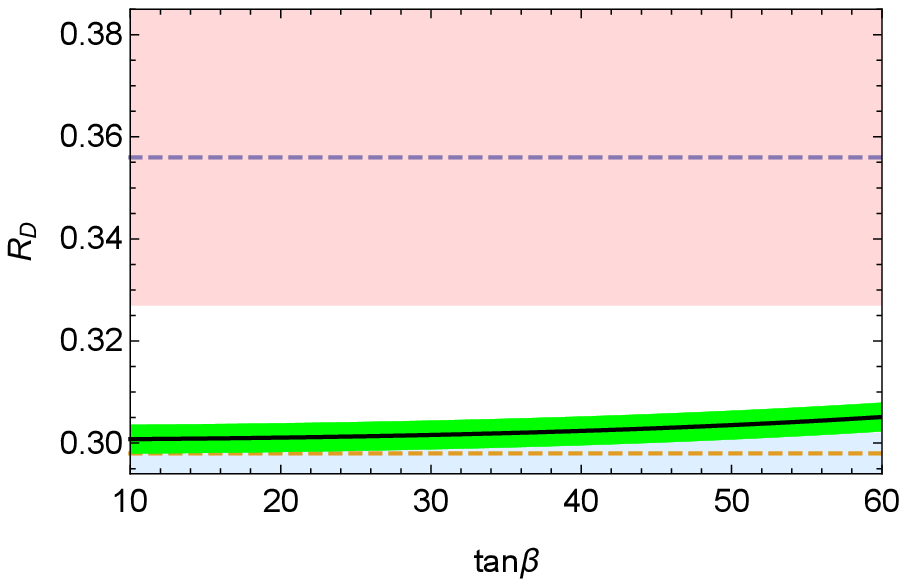}
\end{minipage}
\begin{minipage}[c]{0.48\textwidth}
\includegraphics[width=2.8in]{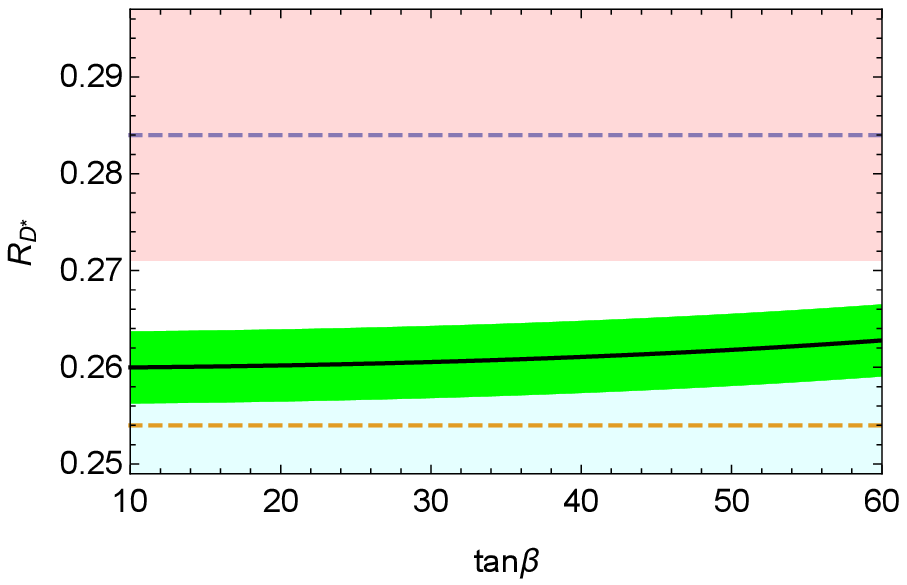}
\end{minipage}
\caption[]{\label{1WTB} The diagrams of $R_{D}$ (left) and $R_{D^{*}}$ (right) versus $\tan\beta$.}
\end{center}
\end{figure}

Similarly, we analyze the effect of the parameter $\tan\beta$ on $R_{D}$ and $R_{D^{*}}$. The parameter $\tan\beta$ is ratio of the VEVs of the two Higgs doublets ($\tan\beta=\upsilon_{u} / \upsilon_{d}$). As a crucial parameter, $\tan\beta$ can affect the masses of particles by directly affecting $v_u$ and $v_d$, and it almost appears in the mass matrixes of all particles. So it should be an important parameter. Supposing the parameter  $g_{YX}=0.4$, the more reasonable parameter space can be obtained when $\tan\beta$ varies from 10 to 60. Fig.~\ref{1WTB} displays $R_{D}$ (left) and $R_{D^{*}}$ (right) versus $\tan\beta$.

It is easy to see that the numerical results of $R_{D}$ and $R_{D^*}$ are increasing functions of $\tan\beta$  and the change of $R_{D}$ is more obvious. When $\tan\beta$ is $60$, the central value of $R_{D}$ can reach $0.3051$, and the central value of $R_{D^*}$ can reach $0.262$. These central values have also exceeded the range of $\sigma_{SM}$.
When we consider that the errors of $R_{D}$ and $R_{D^*}$ are $0.0028$ and $0.0037$, the biggest value of $R_{D}$ can reach $0.3079$. It can be seen that $U(1)_X$SSM has a positive effect on the numerical correction.

\begin{figure}[!htbp]
\begin{center}
\begin{minipage}[c]{0.48\textwidth}
\includegraphics[width=2.8in]{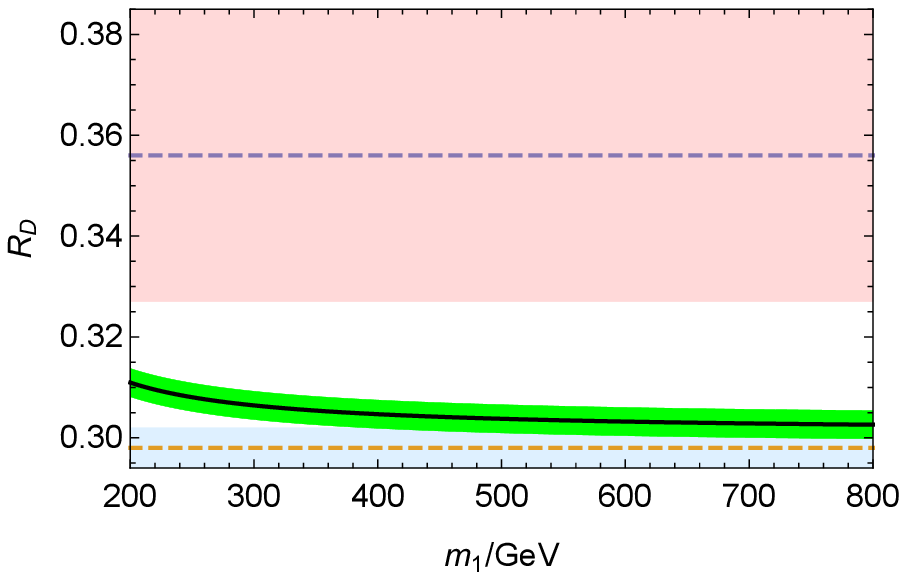}
\end{minipage}
\begin{minipage}[c]{0.48\textwidth}
\includegraphics[width=2.8in]{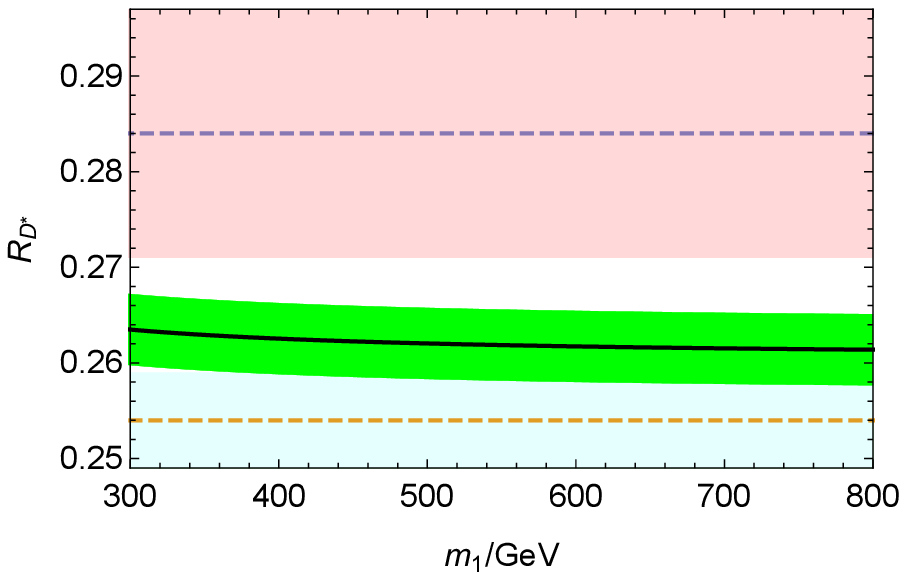}
\end{minipage}
\caption[]{\label{Fig10} The diagrams of $R_{D}$ (left) and $R_{D^{*}}$ (right) versus $m_1$.}
\end{center}
\end{figure}
Then, we set $g_{YX}=0.4$ and $\tan\beta=57$, and plot the $R_{D}$ (left) and $R_{D^{*}}$ (right) varying with $m_1$ in Fig.~\ref{Fig10}. $m_1$ is the $U(1)_Y$ gaugino mass, that affects the mass matrix of the neutralino.
The mass matrix of neutralino in $U(1)_X$SSM is extended to 8$\times$8.
The two lines are all decreasing functions as $m_1$ turns large. The increase of $m_1$ will lead to the increase of neutralino mass,
which suppresses the numerical results. For central value, the biggest and smallest values of $R_{D}$ are 0.3065 and 0.3026, respectively. The values of the $R_{D^{*}}$ vary from 0.263 to 0.261. The value of $R_{D}$ is raised to the range of $2\sigma_{SM}$, and the value of $R_{D^{*}}$ is also above the range of $\sigma_{SM}$. Obviously, $m_1$ has a great influence on the observable measurement.

It is regrettable that, under the influence of the parameters, the theoretical values of $R_{D}$ and $R_{D^{*}}$ still fail to reach the one $\sigma_{exp}$ range of the experimental values, but the predictive values  have improved.

\subsection{The two-dimension scatter plot}
In order to better study the influence of parameters, we draw some multidimensional scatter plots according
to $\chi^2$, that is a statistically sound approach to the study of multidimensional parameter space.
We use the simplified expression of $\chi^2$ as
\begin{eqnarray}
\chi^2=\sum_{i}\left(\frac{\mu_i^{th}-\mu_{i}^{exp}}{\delta_i}\right)^2. \label{kafang}
\end{eqnarray}
In Eq.({\ref{kafang}}), $\mu^{th}_i$ represents the theoretical value for the corresponding procedure which is obtained in $U(1)_X$SSM.
The experimental data is denoted by $\mu^{exp}_i$, and $\delta_i$ represents the error including statistic and system.

The concrete form of $\chi^2$  is shown as
\begin{eqnarray}
\chi^2&&=\left(\frac{m_{h^0}^{th}-m_{h^0}^{exp}}{\delta_{m_{h^0}}}\right)^2+
\left(\frac{\mu_{\gamma\gamma}^{th}-\mu_{\gamma\gamma}^{exp}}{\delta_{\gamma\gamma}}\right)^2+
\left(\frac{\mu_{ZZ}^{th}-\mu_{ZZ}^{exp}}{\delta_{ZZ}}\right)^2
\nonumber\\&&+\left(\frac{\mu_{WW}^{th}-\mu_{WW}^{exp}}{\delta_{WW}}\right)^2
+\left(\frac{\mu_{b\bar{b}}^{th}-\mu_{b\bar{b}}^{exp}}{\delta_{b\bar{b}}}\right)^2
+\left(\frac{\mu_{\tau\bar{\tau}}^{th}-\mu_{\tau\bar{\tau}}^{exp}}{\delta_{\tau\bar{\tau}}}\right)^2
\nonumber\\&&+\left(\frac{R_D^{th}-R_D^{exp}}{\delta_{R_D}}\right)^2
+\left(\frac{R_{D^*}^{th}-R_{D^*}^{exp}}{\delta_{R_{D^*}}}\right)^2
+\left(\frac{\Delta a_\mu^{th}-\Delta a_\mu}{\delta_{\Delta a_\mu}}\right)^2.
\end{eqnarray}
The values of $R_{D}^{exp}$ and $R_{D^*}^{exp}$ are shown in the Eq.(\ref{tnph1}) of the introduction.
The averaged values for the experimental data are adopted from the updated PDG\cite{pdg} in 2023,
$m_{h^0}^{exp}=125.25\pm0.17{\rm GeV},~\mu_{\gamma\gamma}^{exp}=1.10\pm0.07,
~\mu_{ZZ}^{exp}=1.02\pm0.08,~\mu_{WW}^{exp}=1.00\pm0.08,~\mu_{b\bar{b}}^{exp}=0.99\pm0.12,
~\mu_{\tau\bar{\tau}}^{exp}=0.91\pm0.09$.
$\Delta a_\mu=(2.51\pm0.59)\times10^{-11}$ is obtained from the latest work of muon g-2 \cite{UX3,BNFN}.

\begin{figure}[!htbp]
\begin{center}
\begin{minipage}[c]{0.48\textwidth}
\includegraphics[width=3in]{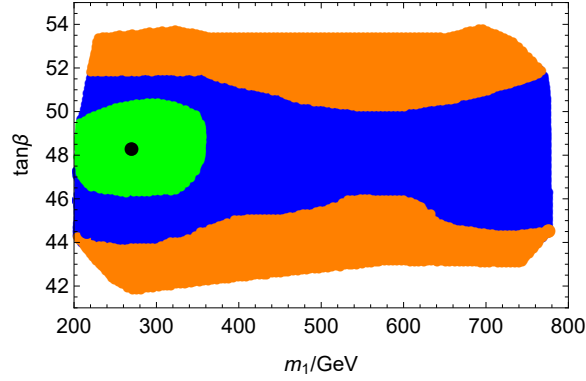}
\end{minipage}
\caption[]{ The $\chi^2$ diagram in the plane of $m_1-\tan\beta$.\label{TBM1}}
\end{center}
\end{figure}

\begin{figure}[!htbp]
\begin{center}
\begin{minipage}[c]{0.48\textwidth}
\includegraphics[width=3in]{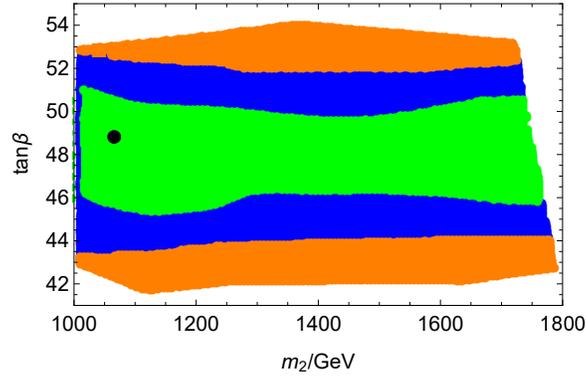}
\end{minipage}
\caption[]{ The $\chi^2$ diagram in the plane of $m_2-\tan\beta$.\label{TBM2}}
\end{center}
\end{figure}

We assume the parameters as $g_X=0.35,~g_{YX}=0.26,~\left(M^2_l\right)_{11} = 0.3~{\rm TeV}^2,~\left(M^2_l\right)_{22} = 0.13~{\rm TeV}^2,~\left(M^2_l\right)_{33} = 0.15~{\rm TeV}^2,~\left(M^2_e\right)_{33} = 4~{\rm TeV}^2$. The $\chi^2$ plots are shown in the Fig. \ref{TBM1} and Fig. \ref{TBM2}.
In the figure, the black circle $\bullet$ represents the best-fitted benchmark point($\chi^2_{min}$).
The green area indicates that $\chi^2\leq\chi^2_{min}+2.28$, which has a confidence level of $68.3\%$.
The confidence level of $93.45\%$ is shown in the blue area, where $\chi^2$ is greater than $\chi^2_{min}+2.28$ and less than $\chi^2_{min}+6.2$.
While the orange area are at the $99.73\%$ confidence level, which are in the region
$\chi^2_{min}+6.2\le\chi^2\leq\chi^2_{min}+11.82$.

We set $m_2 = 1.8~{\rm TeV}$ and draw Fig. \ref{TBM1} with $\tan\beta$ and $m_1$ as variables. $\tan\beta$ is less than $60$ and greater than $40$, and $0.2~ {\rm TeV}\leq m_1\leq0.8~{\rm TeV}$. In this case $\chi^2_{min} = 22.24$. The green area is on the left side of the figure,
where $m_1$ is less than $360~{\rm GeV}$ and $\tan\beta$ is greater than $46$ and less than $50.5$.
In other words, $m_1$ has considerable effect on $\chi^2$. If the value of $m_1$ is small, the value of $\chi^2$ is also small.
Similarly, $\tan\beta$ also has a significant impact on $\chi^2$. The figure is nearly symmetrical in the horizontal direction.

In Fig. \ref{TBM2}, we take $\tan\beta$ and $m_2$ as variables. When $m_1 = 290~{\rm GeV}$, the minimum value of $\chi_{min}$ is $21.97$. For the green areas, $\tan\beta$ is between $45$ and $51$. From the middle to both sides, the value of $\chi^2$ gradually rising.
Similar to Fig. \ref{TBM1}, Fig. \ref{TBM2} also exhibits approximate symmetry,
and the degree of symmetry is stronger.
That is to say, $m_2$ has a relatively smaller effect on the results compared with $\tan\beta$.

In Fig. \ref{TBM1}, the minimum value of $\chi^2$ is reached when $\tan\beta=48$ and $m_1=270~{\rm GeV}$.
In Fig. \ref{TBM2}, as $\tan\beta=49$ and $m_2=1060~{\rm GeV}$, $\chi^2$ takes the minimum value.
Combining the two figures, it can be seen that $m_2$ has a comparatively weak influence among the three sensitive parameters.
The parameter with the strongest impact on the results is $\tan\beta$ for three variables.

\subsection{The multidimensional scatter plot}
\begin{figure}[!htbp]
\begin{center}
\begin{minipage}[c]{0.48\textwidth}
\includegraphics[width=3in]{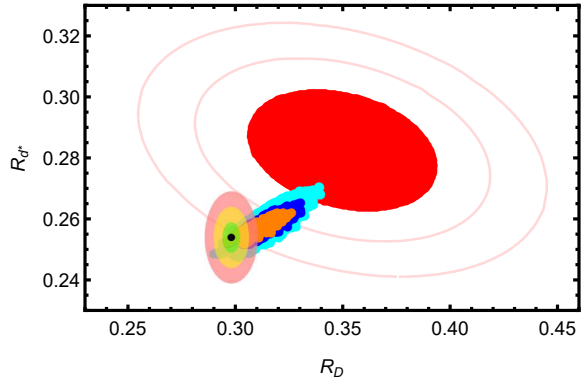}
\end{minipage}
\caption[]{\label{Fig9}
The latest averaged experimental results of $R_D$ and $R_{D^*}$ (red, 1, 2, 3$\sigma$ contours),
compared with our theoretical results(orange, blue, cyan) and the SM
prediction({$\bullet$} represents SM theoretical central value,
transparent green, yellow, red denote the 1, 2, 3$\sigma$)\label{RDRDX}.}
\end{center}
\end{figure}

Through the above figure, we separately show the influence of
parameters that are more sensitive than others on the results. Next,
in order to show the influence of $U(1)_X$SSM on the results more completely,
 we select seven variables at the same time to spread the points with their regions
\begin{eqnarray}
&&100~ {\rm GeV}\leq m_1\leq300~{\rm GeV},~~45\leq\tan{\beta}\leq60,~~
1 ~{\rm TeV}\leq m_2\leq2~{\rm TeV},
\nonumber\\&& 0.1~ {\rm TeV}^2\leq \left(M^2_l\right)_{11}\leq 0.4~{\rm TeV}^2,~~0.1~ {\rm TeV}^2\leq \left(M^2_l\right)_{22}\leq 0.4~{\rm TeV}^2,
\nonumber\\&&0.1~ {\rm TeV}^2\leq \left(M^2_l\right)_{33}\leq 0.4~{\rm TeV}^2,
~~0.3~ {\rm TeV}^2\leq \left(M^2_e\right)_{33}\leq 0.5~{\rm TeV}^2.
\end{eqnarray}
At the same time, we set $g_X$=$0.35$ and $g_{YX}$=$0.26$ to get Fig. \ref{RDRDX}.

In this figure, the contour plot of $R_D-R_D^*$ for the
latest averaged experimental results (red, 1, 2, 3$\sigma$) is shown.
The SM theoretical central value is plotted by the {$\bullet$}.
While, the 1, 2, 3 $\sigma$ of SM prediction are marked by transparent green, yellow and red.
In the figure, the obtained numerical results in our model are divided into three ranges of orange, blue and cyan based on confidence 68.3\%, 95.45\%,~99.73\%. Obviously, the SM central value is out of the 3$\sigma$ contour of the experiment values of $R_D$ and $R_{D^*}$.
Considering 3$\sigma$ effect of SM prediction, the SM prediction can enter the 3$\sigma$ contour of the
experimental value and almost reach experimental 2$\sigma$ contour.
In the whole, the departure between SM prediction and experiment value is obvious.
The obtained numerical results overlap the SM central value,
and stretch over 1, 2, 3$\sigma$ region of the experimental value.
Most points are concentrated in the 2, 3$\sigma$ strips.
At the same time, some obtained numerical results are outside of 3$\sigma_{exp}$ contour and
some obtained numerical results are inside of 1$\sigma_{exp}$ contour.
Considering multiple experiment constraints, the numerical results are restricted tightly
in the $\chi^2$ analysis. At $95.45\%$ confidence level,
our numerical results barely reach the edge of 1$\sigma_{exp}$. When
taking into account values above $95.45\%$ confidence level, the numerical results enter the 1$\sigma_{exp}$ range.
Then our results do not reach the experiment central value.
Despite this, our results are still much better than the SM prediction.

\section{discussion and conclusion}
The departure between the SM prediction of $R_D$ and the actual experimental
value is $2\sigma_{exp}$. The corresponding deviation
of $R_{D^*}$ reaches $2.3\sigma_{exp}$.
Therefore, we will consider the new physics effect and reduce the particle mass in other models to correct the theoretical value of B meson decay. The local gauge group of $U(1)_X$SSM is $SU(3)_C\otimes SU(2)_L \otimes U(1)_Y\otimes U(1)_X$, which makes it to contain new superfields such as right-handed neutrinos $\hat{\nu}_i$ and three Higgs superfields $\hat{\eta},~\hat{\bar{\eta}},~\hat{S}$. In the model, the influence of the right-handed neutrinos and three Higgs superfields is introduced on the mass matrixes, as well as the new influence of the two Abelian groups. These have an impact on the correction of the theoretical value.

We obtain the Wilson coefficients by amplitude calculation, and further calculate the observables. Through numerical analysis, we find that $R_{D^{(*)}}$ is more dependent on $\tan\beta$, $m_1$, $m_2$ and $g_{YX}$.
Supposing several parameters as variables,
we use $\chi^2$ analysis and take into account multiple experiment constraints including
the 125.25 GeV Higgs mass and its decays $h^0\rightarrow \gamma\gamma,~ ZZ,~ WW,~ b\bar{b},~\tau\bar{\tau}$.
The anomaly of muon g-2 is also of interest, and it is considered here.
Our numerical results span in 1, 2, 3$\sigma$ region of the experimental value.
There are many obtained numerical results in the 2, 3$\sigma_{exp}$ strips. Considering results above $95.45\%$ confidence level, our values can appear in the 1$\sigma_{exp}$ range.
Although our results of $R_D$ and $R_{D^*}$ do not arrive at the experiment central values,
they are greatly improved compared with the SM theoretical predictions. In the previous work on BLMSSM, results of $R_D$ and $R_{D^*}$ have been improved. Under the $U(1)_X$SSM,
the theoretical predictions deserve to be further improved.
In other words, NP plays a positive role in increasing the theoretical predicted values of $R_{D^{(*)}}$.
The experimental values also decrease gradually year by year. It is believed that with the efforts of the experimental group and the continuous research on NP, the gap between theoretical and experimental values is expected to be smaller in the future.

{\bf Acknowledgments}

This work is supported by National Natural Science Foundation of China (NNSFC)
(No. 12075074), Natural Science Foundation of Hebei Province
(A2020201002, A202201022, A2022201017), Natural Science Foundation of Hebei Education
Department (QN2022173), Post-graduate's Innovation Fund Project of Hebei University
(HBU2023SS043), the youth top-notch talent support program of the Hebei Province.

%%%%%%%%%%%%%%%%%%%%%%%%%%%%%%%%%%%%%%%%%%%%%
%\clearpage
\appendix
\begin{center}
\Large{{\bf Appendix}}
\end{center}

\section{Full expressions for $a_\ell^{D^*}$ and $c_\ell^{D^*}$ }\label{alcl}

The expressions for $a_\ell^{D^*}$ and $c_\ell^{D^*}$ are shown here.
{\small \begin{eqnarray}
&&a_\ell^{D^\ast} (-) = \frac{8 M_B^2 \pds}{\mbmd^2} {\bf \cvlsq \vsq} +
\frac{\mbmd^2 ( 8 M_{D^*}^2 q^2 + \lambda)}{2 M_{D^*}^2 q^2} {\bf \calsq \aosq}
\nonumber \\
&&+\frac{8 M_B^4 |p_{D^\ast}|^4 }{\mds \mbmd ^2 q^2} {\bf \calsq \atsq} -  \frac{4\pds M_B^2 \mbmdqsq }{\mds q^2} {\bf \calsq  A_1 A_2}\nonumber\\
&&{+\frac{ 32 \mB^2 {\pDmag}^2 }{q^2} {\bf {\ctrsq} T_1^2 } + \frac{ 8
\left(\mB^2-\mDs^2\right)^2 }{q^2} {\bf {\ctrsq} T_2^2 }}\nonumber\\
&& + m_\ell \left[\frac{32 M_B^2 \pds}{q^2 \mbmd} {\bf \re\left(\cvl \ctlconj+\cvr \ctrconj \right)
V T_1} \nonumber \right. \\
&& \left. + {\bf\re\left(\call \ctlconj \right)}\Big(\frac{8 \mbmd \left(2 \mds \mbmdsq + M_B^2 \pds \right)}{q^2 \mds}
{\bf A_1 T_2} \nonumber \right. \\
&&\left. - \frac{8 M_B^2 \mbmdqsq \pds}{q^2 \left(M_B - M_{D^\ast}\right) \mds}
{\bf  A_1 T_3} \nonumber \right.  \left. - \frac{8 M_B^2 \mbmdqthsq \pds}{q^2 \mbmd \mds} {\bf  A_2 T_2} \right. \nonumber \\
&& \left. + \frac{32 M_B^4 |p_{D^\ast}|^4}{q^2 \mds \mbmd \mbmdsq} {\bf  A_2 T_3} \right.\Big)\nonumber\\
&&  -\frac{8 (\mB-\mDs) (\mB+\mDs)^2 }{q^2} {\bf
\re\left(\car \ctrconj\right) A_1 T_2} \bigg]\nonumber\\
&&+ m_\ell^2 \left[\frac{32 M_B^2 \pds}{q^4} {\bf \ctlsq T_1^2}  \right. \left. + \frac{32 M_B^4 |p_{D^\ast}|^4}{q^2 M_{D^\ast}^2 \mbmdsq^2}
{\bf \ctlsq T_3^2}\right.\nonumber \\
&& \left. + \frac{2 \left(8 \mds\left(2 \left(M_B^2 + M_{D^\ast}^2\right) -
q^2\right) q^2 + \left(4 \mds + q^2\right)\lambda\right)}{q^4 \mds} {\bf \ctlsq
T_2^2} \nonumber \right. \nonumber\\
&& \left.- \frac{16 M_B^2 \pds \mbmdqthsq}{q^2 \mds \mbmdsq} {\bf
\ctlsq T_2 T_3} \right].
\end{eqnarray}}
{\small\begin{eqnarray}
&&a_\ell^{D^\ast} (+) = \frac{8 \pds M_B^2}{\mbmc^2} {\bf \cplsq \azsq} +
\frac{32 M_B^2 \pds}{q^2} {\bf \ctlsq T_1^2} + \frac{8 \mbmdsq^2}{q^2} {\bf
\ctlsq T_2^2} \nn \\
&&- m_\ell \bigg[\frac{16 \pds M_B^2}{\mbmc q^2}{\bf \re \left(\call \cplconj
\right)\azsq}   - \frac{32 M_B^2 \pds}{q^2 \mbmd} {\bf \re\left(\cvl \ctlconj+\cvr\ctrconj\right) V
T_1} \nonumber\\
&&  - \frac{8(M_B + M_{D^\ast})\mbmdsq}{q^2}{\bf \re \left(\call
\ctlconj\right) A_1 T_2} \nn\\
&& + {\bf\re\left(\car \ctrconj\right)}\Big(\frac{8 (\mB+\mDs) \left(-2 \mDs^4+\mB^2
\left(2 \mDs^2+\pDmag^2\right)\right) }{\mDs^2 q^2}{\bf
A_1 T_2} \nn\\
&&+\frac{ 8 \mB^2 {\pDmag}^2 \left(-\mB^2+\mDs^2+q^2\right)
}{(\mB-\mDs) \mDs^2 q^2} {\bf  A_1 T_3} -\frac{ 8 \mB^2 \pDmag^2 \left(\mB^2+3
\mDs^2-q^2\right) }{\mDs^2 (\mB+\mDs) q^2} {\bf  A_2
T_2}\nn\\
&&{+\frac{32\mB^4 \pDmag^4 }{(\mB-\mDs)
\mDs^2 (\mB+\mDs)^2 q^2} {\bf  A_2 T_3}}\Big)\bigg]\nn\\
&& + m_\ell^2 \left[\frac{8 \pds M_B^2}{q^4} {\bf \calsq \azsq}
+\frac{8 \pds M_B^2}{\mbmd^2 q^2} {\bf \cvlsq \vsq} \right. \left. + \frac{2 \mbmd^2}{q^2} {\bf \calsq \aosq} \right.\nn\\
&&{+{\bf \ctrsq}\Big(\frac{32 \mB^2 {\pDmag}^2 }{q^4} {\bf T_1^2} + 8 \frac{
 \mB^2
\pDmag^2 }{\mDs^2 q^2} {\bf T_2^2} }{+\frac{16 \left(\mB^4+\mDs^4-2 \mB^2
\left(\mDs^2+\pDmag^2\right)\right)}{q^4} {\bf T_2^2}}\nn\\
&&{+\frac{32 \mB^4 \pDmag^4 }{\left(-\mB^2
\mDs+\mDs^3\right)^2 q^2} {\bf T_3^2}}{+\frac{16  \mB^2 \pDmag^2 \left(\mB^2+3
\mDs^2-q^2\right) }{\mDs^2 \left(-\mB^2+\mDs^2\right) q^2}
{\bf T_2 T_3}\Big)}\bigg].
\end{eqnarray}}
{\small\begin{eqnarray}
&&c_\ell^{D^\ast} (-) = \frac{8 \pds M_B^2}{\mbmd^2} {\bf{\cvlsq \vsq}} -
\frac{\mbmd^2 \lambda}{2 \mds q^2} {\bf \calsq \aosq} \nn\\
&& - \frac{8 |p_{D^\ast}|^4 M_B^4}{\mbmd^2 \mds q^2} {\bf \calsq \atsq} + \frac{4 \pds M_B^2 \mbmdqsq}{\mds q^2} {\bf \calsq A_1 A_2} \nn \\
&&{-\frac{32\mB^2 \mDs^2 \left(\mB^2-\mDs^2\right)^2
{\pDmag}^2}{\left(-\mB^2 \mDs+ \mDs^3\right)^2 q^2} {\bf
{\ctrsq}  T_1^2}}{+\frac{2  \left(\mB^2-\mDs^2\right)^2
}{\mDs^2} {\bf {\ctrsq} T_2^2 }}\nn\\ && {-\frac{4
\left(-\mB^2+\mDs^2\right)
\left(-\mB^4+\mDs^4+4 \mB^2 \pDmag^2\right) }{\mDs^2 q^2}
{\bf {\ctrsq} T_2^2}}\nn\\
&&{+\frac{ 32 \mB^4 {\pDmag}^4 }{\left(-\mB^2
\mDs + \mDs^3\right)^2} {\bf {\ctrsq} T_3^2}}{+\frac{ 16 \mB^2 {\pDmag}^2 \left(\mB^2+3
\mDs^2-q^2\right) }{-\mB^2 \mDs^2+\mDs^4} {\bf {\ctrsq} T_2 T_3}}\nn \\
&& + m_\ell \left[\frac{32 M_B^2 \pds}{q^2 \mbmd} {\bf \re\left(\cvl \ctlconj-\cvr\ctrconj\right)
V T_1}  \nn \right. \\
&& +{\bf\re\left(\call \ctlconj \right)}\Big(\left. -\frac{8 M_B^2 \left(M_B + M_{D^\ast}\right)\pds }{q^2 \mds}{\bf
A_1 T_2} \right. \left.+ \frac{8 M_B^2 \mbmdqsq \pds}{q^2 \mds \left(M_B - M_{D^\ast}\right) }
{\bf  A_1 T_3 } \nn \right. \\
&& \left. + \frac{8 M_B^2 \mbmdqthsq \pds}{q^2 \mds \mbmd }{\bf  A_2 T_2} \nn \right.  \left. - \frac{32 M_B^4 |p_{D^\ast}|^4}{q^2 \mds \mbmd \mbmdsq} {\bf
 A_2 T_3} \right.\Big) \nn\\
&& {+\frac{ 8 \mB^2 (\mB+\mDs) {\pDmag}^2 }{\mDs^2
q^2}{\bf \re\left(\car\ctrconj \right) \bf {A_1 T_2}}}\nn\\
&&+{{\bf\re\left(\cvr\ctrconj \right)}\Big(-\frac{ 8 \mB^2 {\pDmag}^2
\left(-\mB^2+\mDs^2+q^2\right) }{(\mB-\mDs) \mDs^2 q^2} {\bf A_1 T_3}}  \nn\\
&&{+\frac{8 \mB^2 {\pDmag}^2 \left(\mB^2+3
\mDs^2-q^2\right) }{\mDs^2 (\mB+\mDs) q^2} {\bf A_2
T_2}}{-\frac{ 32 \mB^4 {\pDmag}^4 }{(\mB-\mDs)
\mDs^2 (\mB+\mDs)^2 q^2} {\bf A_2 T_3}\Big)}\bigg]\nn\\
&& + m_\ell^2 \left[\frac{32 M_B^2 \pds}{q^4} {\bf \ctlsq T_1^2} + \frac{2
\left(4\mds - q^2\right)\lambda}{\mds q^4} {\bf \ctlsq T_2^2} \nn \right. \\
&& \left. - \frac{32 M_B^4 |p_{D^\ast}|^4 }{q^2 \mds \mbmdsq^2} {\bf \ctlsq
T_3^2} + \frac{16 M_B^2 \pds \mbmdqthsq}{q^2 \mds \mbmdsq} {\bf \ctlsq
T_2 T_3} \right].%\\
\end{eqnarray}}
{\small\begin{eqnarray}
&&c_\ell^{D^\ast} (+) = {\bf \ctlsq} \Big(-\frac{32 M_B^2 \pds}{q^2} {\bf T_1^2} - \frac{2
\left(4 \mds - q^2\right)\lambda}{\mds q^2} {\bf  T_2^2} \nn \\
&&+ \frac{32 M_B^4 |p_{D^\ast}|^4 }{\mds \mbmdsq^2 } {\bf  T_3^2}
- \frac{16 M_B^2 \pds \mbmdqthsq}{\mds \mbmdsq} {\bf T_2 T_3}\Big) \nn
\\
&& - m_\ell \left[\frac{32 M_B^2 \pds}{q^2 \mbmd} {\bf \re\left(\cvl \ctlconj+\cvr\ctrconj\right)
V T_1}  \nn \right. \\
&& +{\bf\re\left(\call \ctlconj\right)}\Big(\left. -\frac{8 M_B^2 \left(M_B + M_{D^\ast}\right)\pds }{q^2 \mds}{\bf
A_1 T_2} \right. \left.+ \frac{8 M_B^2 \mbmdqsq \pds}{q^2 \mds \left(M_B - M_{D^\ast}\right) }
{\bf  A_1 T_3 } \nn \right. \\
&& \left. + \frac{8 M_B^2 \mbmdqthsq \pds}{q^2 \mds \mbmd }{\bf A_2 T_2} \nn \right. \left. - \frac{32 M_B^4 |p_{D^\ast}|^4}{q^2 \mds \mbmd \mbmdsq} {\bf
 A_2 T_3} \right. \Big)\nn  \nn\\
&& + {\bf\re\left(\car \ctrconj\right)}\Big(\frac{ 8 \mB^2 (\mB+\mDs) \pDmag^2 }{\mDs^2
q^2}{\bf A_1 T_2}- \frac{8 \mB^2 \pDmag^2
\left(-\mB^2+\mDs^2+q^2\right) }{(\mB-\mDs) \mDs^2 q^2} {\bf
 A_1 T_3}  \nn\\
&&+\frac{8\mB^2 \pDmag^2 \left(\mB^2+3
\mDs^2-q^2\right) }{\mDs^2 (\mB+\mDs) q^2} {\bf A_2
T_2}- \frac{32  \mB^4 \pDmag^4 }{(\mB-\mDs)
\mDs^2 (\mB+\mDs)^2 q^2} {\bf A_2 T_3}\Big)\bigg]\nn\\
&& + m_\ell^2 \left[ - \frac{8 \pds M_B^2}{\mbmd^2 q^2} {\bf \cvlsq \vsq} +
\frac{\mbmd^2 \lambda}{2 \mds q^4} {\bf \calsq \aosq} \right. \nn\\
&& \left. + \frac{8 |p_{D^\ast}|^4 M_B^4}{\mds \mbmd^2 q^4} {\bf \calsq \atsq}
\right.\left.- \frac{4 \pds M_B^2}{\mds q^4} \mbmdqsq {\bf \calsq A_1
A_2}\right.\nn\\
&&\left. +\frac{ 32 \mB^2 \pDmag^2 }{q^4} {\bf \ctrsq T_1^2 } +\frac{8 \mB^2 \pDmag^2 \left(4 \mDs^2-q^2\right)
}{\mDs^2 q^4} {\bf  \ctrsq T_2^2 }\right. \nn\\
&&\left.-\frac{32  \mB^4 \pDmag^4 }{\left(-\mB^2 \mDs+\mDs^3\right)^2 q^2} {\bf \ctrsq T_3^2}\right.\left.+\frac{16  \mB^2 \pDmag^2 \left(\mB^2+3 \mDs^2-q^2\right) }{\mDs^2 \left(\mB^2-\mDs^2\right) q^2} {\bf \ctrsq T_2 T_3}\right].
\end{eqnarray}}

The form factors $\textbf{V}, \textbf{A}_0, \textbf{A}_1, \textbf{A}_2,
\textbf{T}_1, \textbf{T}_2$ and $\textbf{T}_3$  are not calculated by Lattice QCD. Here we use the heavy quark effective theory (HQET) form factors. The specific forms are as follows\cite{36,alcl2},

\begin{equation}
\begin{split}
&{\bf V} = \frac{1 + r}{2 \sqrt{r}} h_V,\\
&{\bf A_1} = \frac{\sqrt{r}\,(1 + w)}{1 + r} h_{A_1},\\
&{\bf A_2} = \frac{1+r}{2\sqrt{r}}\left(r h_{A_2} + h_{A_3}\right),\\
&{\bf A_0} = \frac{1}{2 \sqrt{r}} \left[ (w + 1) h_{A_1} + (r w - 1) h_{A_2} + (r - w) h_{A_3}\right],\\
&{\bf T_1} = -\frac{1}{2\sqrt{r}}\left[(1 - r) h_{T_2} - (1 + r) h_{T_1}\right],\\
&{\bf T_2} = \frac{1}{2\sqrt{r}}\left[\frac{2 r (w + 1)}{1 + r} h_{T_1}-\frac{2r(w - 1)}{1 - r} h_{T_2}\right],\\
&{\bf T_3} = \frac{1}{2\sqrt{r}}\left[(1 - r) h_{T_1} - (1 + r) h_{T_2} + \left(1 - r^2\right) h_{T_3} \right].\\
\end{split}
\end{equation}
where,
\begin{align}
      \begin{split}
         h_V(w) =& R_1(w) h_{A_1}(w), \\
         h_{A_2}(w) =& \frac{R_2(w)-R_3(w)}{2\,r_\Dst } h_{A_1}(w), \\
         h_{A_3}(w) =& \frac{R_2(w)+R_3(w)}{2 } h_{A_1}(w), \\
         h_{T_1}(w) =& \frac{ 1} {2 ( 1 + r_\Dst^2 - 2r_\Dst w ) } \left[ \frac{ m_b - m_c }{M_B - M_\Dst } ( 1 - r_\Dst )^2 ( w + 1 ) \, h_{A_1}(w) \right. \\
                     & \quad\quad\quad\quad\quad\quad\quad\quad\quad \left. - \frac{ m_b + m_c }{ M_B + M_\Dst } ( 1 + r_\Dst )^2 ( w - 1 ) \, h_V(w) \right],  \\
         h_{T_2}(w) =& \frac{ ( 1 - r_\Dst^2 ) ( w + 1 ) }{2 ( 1 + r_\Dst^2 - 2r_\Dst w ) } \left[ \frac{ m_b - m_c }{ M_B - M_\Dst } \, h_{A_1}(w) -
         \frac{ m_b + m_c }{ M_B + M_\Dst } \, h_V(w) \right],
      \end{split}
\end{align}
\begin{align}
      \begin{split}
         h_{T_3}(w) =& -\frac{ 1 }{2 ( 1 + r_\Dst ) ( 1 + r_\Dst^2 - 2r_\Dst w ) } \left[2 \frac{ m_b - m_c }{ M_B - M_\Dst } r_\Dst ( w + 1 ) \, h_{A_1}(w) \right. \\
                     & - \frac{ m_b - m_c }{ M_B - M_\Dst } ( 1 + r_\Dst^2 - 2r_\Dst w ) ( h_{A_3}(w) - r_\Dst h_{A_2}(w) ) \\
                     & \left. - \frac{ m_b + m_c }{ M_B + M_\Dst } ( 1 + r_\Dst )^2 \, h_V(w) \right],
      \end{split}
\end{align}
\begin{align}
   \begin{split}
      h_{A_1}(w) =& h_{A_1}(1) [ 1 - 8\rho_\Dst^2 z + (53\rho_\Dst^2-15) z^2 - (231\rho_\Dst^2-91) z^3 ],  \\
      R_1(w) =& R_1(1) - 0.12(w-1) + 0.05(w-1)^2,  \\
      R_2(w) =& R_2(1) + 0.11(w-1) - 0.06(w-1)^2,  \\
      R_3(w) =& 1.22 - 0.052(w-1) + 0.026(w-1)^2.
   \end{split}
\end{align}
Here, $r_{D^\ast} = M_{D^\ast}/M_B$, $w(q^2)=(M_B^2+M_{D^{*}}^2-q^2)/(2M_B M_{D^{*}})$ and
$z(w) = ( \sqrt{w+1} - \sqrt2 ) / ( \sqrt{w+1} + \sqrt2 )$.

\end{document}